\documentclass[sn-basic,Numbered]{sn-jnl}

\usepackage{graphicx}%
\usepackage{multirow}%
\usepackage{amsmath,amssymb,amsfonts}%
\usepackage{amsthm}%
\usepackage{mathrsfs}%
\usepackage[title]{appendix}%
\usepackage{xcolor}%
\usepackage{textcomp}%
\usepackage{manyfoot}%
\usepackage{booktabs}%
\usepackage{tabularx}%
\usepackage{algorithm}%
\usepackage{algorithmicx}%
\usepackage{algpseudocode}%
\usepackage{listings}%
\usepackage{tcolorbox}
\usepackage{subcaption} 
\usepackage{pgffor}   
\usepackage{array}
\usepackage[figuresright]{rotating}
\usepackage{pdflscape}                  
\usepackage{siunitx}                    
\usepackage{csquotes}

\begin{document}

\title[Article Title]{Enhancing Requirements Traceability Link Recovery: A Novel Approach with T-SimCSE}

\author[1,2]{\fnm{Ye} \sur{Wang}}\email{yewang@zjgsu.edu.cn}

\author[1,2]{\fnm{Wenqing} \sur{Wang}}\email{23020100051@pop.zjgsu.edu.cn}

\author[1,2]{\fnm{Kun} \sur{Hu}}\email{21020100080@pop.zjgsu.edu.cn}

\author[1,2]{\fnm{Qiao} \sur{Huang}}\email{qiaohuang@zjgsu.edu.cn}

\author[3]{\fnm{Liping} \sur{Zhao}}\email{liping.zhao@manchester.ac.uk}

\affil[1]{\orgdiv{School of Computer Science and Technology}, \orgname{Zhejiang Gongshang University}, \orgaddress{\state{ZheJiang},\city{HangZhou},\postcode{310018}, \country{China}}}

\affil[2]{\orgname{Zhejiang Key Laboratory of Big Data and Future E-Commerce Technology}, \orgaddress{\state{ZheJiang},\country{China}}}

\affil[3]{\orgdiv{Department of Computer Science}, \orgname{The University of Manchester}, \orgaddress{\city{Manchester}, \postcode{M13 9PL}, \country{United Kingdom}}}

\abstract{Requirements traceability plays an important role in ensuring software quality and responding to changes in requirements. Requirements trace links (such as the links between requirements and other software artifacts) underpin the modeling and implementation of requirements traceability. With the rapid development of artificial intelligence, more and more pre-trained language models (PLMs) techniques are applied to the automatic recovery of requirements trace links. However, the requirements traceability links recovered by these approaches are not accurate enough, and many approaches require a large labeled dataset for training. Currently, there are very few labeled datasets available. To address these limitations, this paper proposes a novel requirements traceability link recovery approach called T-SimCSE, which is based on a PLM---SimCSE. SimCSE has the advantages of not requiring labeled data, having broad applicability, and performing well. T-SimCSE firstly uses the SimCSE model to calculate the similarity between requirements and target artifacts, and  employs a new metric (i.e. specificity)
to reorder those target artifacts. Finally, the trace links are created between the requirement and the top-K target artifacts. We have evaluated T-SimCSE on ten public datasets by comparing them with other approaches. The results show that T-SimCSE achieves superior performance in terms of recall and Mean Average Precision (MAP). }


\maketitle

\section{Introduction}\label{sec1}

Software traceability is defined as \enquote{the ability to interrelate any uniquely identifiable software engineering artifact to any other, maintain required links over time, and use the resulting network to answer questions of both the software product and its development process}~\citet{gotel2012traceability}. Research on software traceability spans multiple dimensions, including traceability recovery, modeling, evaluation, and implementation~\citet{bib2}. Despite its theoretical maturity, the practical adoption of traceability remains hindered by challenges related to scalability and accuracy.

Traceability plays a critical role in ensuring software quality and maintenance efficiency. It enables clear connections among various development stages—requirements, design, implementation, and testing—ensuring the correct implementation of functionalities and manageable change impacts. Moreover, it facilitates efficient debugging, supports collaboration, satisfies compliance requirements, and promotes knowledge accumulation and reuse, thereby enhancing the overall manageability of the software lifecycle.

While existing studies have extensively investigated requirement-to-code traceability, recovering links between requirements and natural language (NL) artifacts —--such as user cases, design documents—--poses distinct challenges. Unlike code artifacts containing explicit structural cues (e.g., call dependencies, data flows)~\citet{bib6, bib7}, NL artifacts often exhibit linguistic heterogeneity due to inconsistent writing conventions, terminological discrepancies, and semantic ambiguities in natural language. For example, a requirement stating \enquote{the system shall encrypt user data} might link to a test case referencing \enquote{secure data transmission} without explicit lexical overlap, confounding traditional keyword-based approaches. This fundamental disparity motivates our focused investigation on requirement-to-NL artifact link recovery. Research~\citet{bib8, guo2017semantically} has shown that building and maintaining trace linkis a labor-intensive, arduous, and error-prone task when dealing with large-scale, complex systems. Researchers have proposed a number of semi-automated approaches to reduce manual operations. Among them, IR (Information Retrieval)-based approaches are the most widely used, which calculates the text similarity between two artifacts by IR model, like Vector Space Model (VSM)~\citet{antoniol2002recovering}, Latent Semantic Indexing (LSI)~\citet{bib10} and the probabilistic Jensen-Shannon model~\citet{bib11}. A list of trace links is then obtained by sorting the artifacts from the largest to the smallest according to the calculated similarity, with the higher-ranked artifacts being considered more likely to be linked. 

However, the accuracy of IR-based approaches for traceability link recovery is usually not satisfying. The main reason is that the writing standards and terminology of the requirements are different from those of NL artifacts~\citet{bib12}, so IR-based approaches do not go well with word matching. To solve this problem, researchers began experimenting with deep learning (DL) techniques to recover trace links, and they used neural networks to solve the word mismatch problem. In contrast, DL-based approaches achieve higher accuracy and recall rates compared to IR-based approaches. However, DL-based approaches require large amounts of data to train neural network models. There is rarely enough data available for training in real development projects, making it difficult to apply such approaches well in the industry.

To accurately measure the semantic similarity between requirements and other software artifacts, we adopt Simple Contrastive Sentence Embeddings (SimCSE)~\citet{gao2021simcse}, a contrastive learning-based sentence embedding model. Its ability to produce high-quality semantic representations while requiring minimal labeled data makes it particularly suitable for requirements traceability tasks. A detailed discussion of the model and the rationale for its selection is provided in Section~\ref{sec3}.

To solve this problem, we propose T-SimCSE, which combines SimCSE and a rewarding strategy of trace links with different rewarding values. Essentially, this work is also an approach based on IR, combined with the pre-trained language model(PLM) to obtain sentence embeddings. First, SimCSE is chosen as the underlying language model to obtain the semantic similarity of two artifacts by comparing their cosine similarity of sentence embeddings, and a list of target artifacts is obtained by sorting them in descending order according to their semantic similarity to the source artifact. We define the relationship termed specificity between the requirement and other artifacts. The fewer closely related artifacts an artifact has, the higher its specificity is. We establish a rewarding mechanism based on specificity: the higher the specificity, the greater the reward. This means that the artifact will be ranked higher during the rewarding process. As some target artifacts may contain generic semantics and are similar to many artifacts, most established approaches treat all semantically close artifacts of the target artifact as equally important, which is unfair. In T-SimCSE, the target artifacts that contain generic semantics are not given the same rewarding value as other semantically close artifacts, and moreover, their rewarding values will be reduced appropriately. We evaluated the T-SimCSE approach by applying it to ten public datasets and the results showed that it performs better than other approaches on part datasets.

The contributions of this paper are as follows:

1) The use of SimCSE effectively addresses the problem of insufficient training data. The results show that T-SimCSE achieves superior performance compared to prior approaches in terms of $F_1$ and $F_2$ scores on MODIS datasets. And the results show that T-SimCSE outperforms the baselines in terms of both Mean Average Precision (MAP) on ten datasets.

2) The design of the rewarding strategy that can reward target artifacts differently depending on their importance, thereby improving their ranking in the list.

The rest of the paper is organized as follows. Section~\ref{sec2} presents the related work. Section~\ref{sec3} elaborates on the motivation and concepts, including a motivating example, the rationale for selecting SimCSE, and the concepts. Section~\ref{sec4} introduces the proposed methodology, detailing the key components and techniques employed. Section~\ref{sec5} describes the experimental setup, including datasets, evaluation metrics, and implementation details. Section~\ref{sec6} provides an in-depth analysis and discussion of the experimental results. Section~\ref{sec7} discusses the implications of the research. Section~\ref{sec8} addresses the potential threats to the validity of our study. Finally, Section~\ref{sec9} concludes the paper with a summary of contributions and potential future research directions.

\section{Related work}\label{sec2}

In this section, we describe traceability link recovery approaches and PLMs applicable to document embedding.

\subsection{Traceability link recovery approaches}\label{subsec2.1}

\begin{table}[htbp] 
  \centering
  \caption{Summary of traceability link recovery approaches}
  \label{tab1}
  \setlength{\tabcolsep}{4pt} 
  \begin{tabularx}{\linewidth}{@{}>{\raggedright\arraybackslash}p{2.2cm} l c >{\raggedright\arraybackslash}X@{}}
    \toprule
    \textbf{Category} & \textbf{Name} & \textbf{Reference} & \textbf{Types of Artifacts} \\
    \midrule
    \multirow[t]{19}{2.2cm}{\raggedright Approaches based on unsupervised models }
      & - & \citet{antoniol2002recovering}       & Code, Manual, Req\tnote{1} \\
      & - & \citet{mahmoud2015information}      & Code, Req \\
      & - & \citet{marcus2003recovering}        & Code, Manual, Req \\
      & - & \citet{asuncion2010software}        & Code, Architectural Description, Collaborative Artifact, Multimedia Material \\
      & - & \citet{hayes2003improving}           & HR\tnote{2}, LR\tnote{3}, Design element \\
      & RETRO & \citet{hayes2006advancing}          & HR, LR, Design element \\
      & - & \citet{gethers2011integrating}      & UC\tnote{4}, UML Diagram, TC\tnote{5}, Code \\
      & COMET & \citet{moran2020improving}          & Req, UC, Code, TC \\
      & WELR, WQI & \citet{zhao2017improved}        & HR, LR, UC, Code, ID\tnote{6}, TC, class description \\
      & S2Trace & \citet{chen2019enhancing}         & HR, LR, UC, Code, ID, TC, class description \\
      & - & \citet{mahmoud2016detecting}        & Req, Code \\
      & - & \citet{schlutter2021improving}      & HR, LR \\
      & TAROT & \citet{gao2023using}               & Code, Req \\
      & FTLR & \citet{hey2021improving}            & Code, UC, Req \\
      & GeT2Trace & \citet{chen2021}                & HR, LR, Code, UC, TC, ID \\
      & - & \citet{liu2020towards}              & Code, Design Definition \\
      & - & \citet{rodriguez2023prompts}        & HR, LR, Req, Code, Design Description \\
       & LiSSA & \citet{fuchss2025lissa}       & Code, Req, Architecture Model, Architecture Documentation \\
      & - & \citet{hey2025requirements}          & Req, Regulatory code, HR, LR \\
    \midrule
     \multirow[t]{10}{2.2cm}{\raggedright Approaches trained on supervised models}
      & - & \citet{lohar2013improving}          & Design, Req, Code, UC, TC \\
      & ALCATRAL & \citet{mills2019tracing}          & HR, LR, ID, TC, UC, Req, Code \\
      & - & \citet{rath2018traceability}        & Commit, Issue, Code \\
      & TRAIL & \citet{mill2018automatic}           & HR, LR, ID, TC, UC, Req, Code \\
      & TLR-ELtoR & \citet{marcen2020traceability}    & TC, Model fragment \\
      & - & \citet{wang2018enhancing}           & HR, LR, Req, Code \\
      & - & \citet{guo2017semantically}         & Design Description, Req \\
      & HyLoc & \citet{lam2015combining}            & Bug Report, Code, Buggy file, API documentation \\
      & T-BERT & \citet{lin2021traceability}        & Issue, Discussion, Commit \\
      & NLTrace & \citet{lin2022enhancing}          & Design Description, Req, Regulation \\
    \bottomrule
  \end{tabularx}
  \begin{tablenotes}
    \footnotesize
    \item[~] \textsuperscript{1}Requirement;
    \textsuperscript{2}High-level Requirement;
    \textsuperscript{3}Low-level Requirement;
    \textsuperscript{4}User Case;
    \textsuperscript{5}Test Case;
    \textsuperscript{6}Interaction Diagram;
  \end{tablenotes}
\end{table}

We categorize traceability link recovery approaches into approaches trained on supervised models and approaches based on unsupervised models, depending on whether they are trained with task-specific supervision. Since different approaches target different artifact types for traceability link recovery, we summarize the artifact types addressed by these approaches in Table~\ref{tab1}. 

\textbf{Approaches based on unsupervised models:} Early information retrieval approaches encompassed the Vector Space Model~\citet{antoniol2002recovering, mahmoud2015information}, Latent Semantic Indexing~\citet{marcus2003recovering}, Latent Dirichlet Allocation (LDA)~\citet{asuncion2010software}, and works by Hayes et al.~\citet{hayes2003improving, hayes2006advancing}. These approaches recovered potential trace links by computing textual similarity between documents. However, they faced challenges including semantic discrepancies and loss of contextual information. For instance, VSM and LSI relied on surface features like term frequency and word co-occurrence, demonstrating limited effectiveness when handling semantically equivalent but lexically different terms.

Subsequent research has focused on integrating complementary techniques to overcome the limitations of traditional approaches. Gethers et al.~\citet{gethers2011integrating} combined VSM with probabilistic models and relational topic models, while Moran et al.~\citet{moran2020improving} enhanced accuracy in their COMET tool through machine learning and transitive links. Semantic comprehension capabilities have been significantly improved through word embeddings~\citet{zhao2017improved} and sequential semantics~\citet{chen2019enhancing}, which capture contextual relationships and temporal dependencies in requirement artifacts. Mahmoud et al.~\citet{mahmoud2016detecting} proposed an unsupervised learning-based approach for non-functional requirement checking, classification, and tracing. For large-scale applications, Schlutter et al.~\citet{schlutter2021improving} introduced semantic relation graphs with spreading activation mechanisms, enabling scalable and interpretable trace queries through automated NLP pipelines. Innovations like TAROT~\citet{gao2023using} continue this integration trend by optimizing IR model weights through consensus bigram phrases in requirements and code, demonstrating how hybrid approaches combine linguistic patterns with statistical rigor for synergistic effects. Hey et al.~\citet{hey2021improving} also used word embeddings and word movers distance to reduce the semantic gap between fine-grained requirements and code. Chen Lei et al.~\citet{chen2021} proposed an unsupervised requirements traceability approach based on graph mining and referenced learning, in order to enhance the semantic representation of artifact text. The approach mines semantic features in the software artifact text network and then uses Doc2Vec and similarity models to learn and compare the semantic similarity of artifacts. Since software projects often include technical terms as well as domain-specific terminology, Liu et al.~\citet{liu2020towards} proposed an approach to automatically generate domain-specific conceptual models. Rodriguez et al.~\citet{rodriguez2023prompts} have shown that without any extra pre-training, through carefully designed prompts, Large Language Model (LLM) can effectively predict the trace links between software artifacts and provide reasonable decision explanations.

\textbf{Approaches trained on supervised models:} Mills et al.~\citet{mills2019tracing} proposed an approach based on active learning that significantly reduces the amount of training data required for supervised classification in constructing trace links without affecting the performance of the algorithm. Rath et al.~\citet{rath2018traceability} combined classification problems, the flow of code changes and text-related features to train classifiers of naive Bayes, DT and random forests to generate the missing links.  Marcén et al.~\citet{marcen2020traceability} proposed a new approach called Evolutionary Learning to Rank for Traceability Link Recovery (TLR-ELtoR), which combines evolutionary learning algorithms and machine learning techniques. Data-driven optimization approaches have also emerged, exemplified by Lohar et al.~\citet{lohar2013improving} who automated traceability feature configuration using historical data to reduce manual intervention.  Wang et al.~\citet{wang2018enhancing} proposed an approach based on word embeddings and feedforward neural network (FNN) that extends the IR-based approach using a word pair ranking model and a cluster ranking model. Word embedding techniques have also been used to construct trace links between requirements and code. For domain-specific trace links, Guo et al.~\citet{guo2017semantically} developed a network for generating trace links between artifacts using an Recurrent Neural Network (RNN) model. Lam et al.~\citet{lam2015combining} proposed a model called HyLoc that combines a Deep Neural Network (DNN) with a revised Vector Space Model (rVSM) to solve the problem of word mismatch for defect localization. DNN and rVSM complement each other and can achieve higher accuracy than individual models. IR-based approaches usually do not match words well because of the semantic differences between the required writing standards and terminology and NL artifacts. DL and ML techniques can solve this problem well, but they both require large amounts of data to train and in the absence of datasets they will be much less effective.

 PLMs are trained on a large-scale unlabeled datasets, learning universal language representations or features. These models perform well across a wide range of tasks because they have acquired rich language structures and knowledge. Subsequently, these PLMs can be fine-tuned on a relatively small annotated dataset to adapt to software traceability, thereby enhancing their performance on the specific task. Models based on Transformer, such as BERT and GPT series, have achieved remarkable success in various natural language processing tasks. Therefore, the application of PLMs in creating and maintaining trace links has gradually become a new research direction. Lin et al.~\citet{lin2021traceability} proposed a BERT-based framework for recovering code-to-requirements trace links while using a three-step training strategy to solve the problem of training datasets, but the approach was only tested on Python projects. Lin et al.~\citet{lin2022enhancing} demonstrated that their NLTrace approach improves software requirements traceability through DL and transfer learning techniques. NLTrace leverages a Language Model as its underlying knowledge base and further enhances the performance of tracing tasks through transfer learning techniques. Fuchss et al.~\citet{fuchss2025lissa} and Hey et al.~\citet{hey2025requirements} combined retrieval-augmented generation (RAG) with LLMs to recover requirements traceability links.

\subsection{Pre-trained language models applicable to document embedding}\label{subsec2.2}

In document embedding tasks, many PLMs have been widely adopted and achieved remarkable results. Sentence-BERT was the first to adapt BERT with a Siamese/triplet objective on NLI and STS data, producing sentence-level vectors whose cosine distance closely mirrors semantic similarity and supporting efficient retrieval in a single forward pass~\citet{reimers2019sentencebert}. Building on the encoder–decoder family, Generalizable T5 Retrieval (GTR-T5) transforms the T5 encoder into a dual-encoder and contrastively trains it on large-scale QA pairs and search logs; models from the \enquote{base} to the 11-billion-parameter tier keep a fixed 768-dimensional space while transferring strongly to BEIR and MTEB zero-shot benchmarks~\citet{ni2022gtrt5}. Microsoft’s E5 goes a step further by instruction-tuning a single encoder on one billion \enquote{query: … / doc: …} pairs, enabling the same network to emit intent-aware vectors for both queries and passages and topping the 2023 MTEB open-source leaderboard~\citet{wang2023e5}. For cross-lingual scenarios, LaBSE combines masked-LM, translation language modeling and bi-text contrastive loss to align 109 languages within a shared 768-dimensional space, achieving state-of-the-art performance in cross-language retrieval~\citet{feng2020labse}. Finally, the BGE (BAAI General Embedding) family employs self-distillation with hard-negative mining, releasing small, base and large checkpoints that dominate both MTEB and C-MTEB leaderboards while remaining inference-efficient for English and Chinese corpora~\citet{xu2023bge}.

\section{Motivations and concepts}\label{sec3}

In this section, we will illustrate the motivations behind our approach through an example and introduce a set of concepts used in our approach.

\subsection{A motivating example}\label{subsec3.1}

To recover requirement traceability links, the traditional approach is usually to directly calculate the semantic similarity of the requirement and target artifacts (TAs). However, in practice, some of the correct TAs are often referenced by other correct TAs during the implementation. Their textual content may not be directly related to the requirement, which will result in the missing of such correct indirect links when traditional approaches are applied. If we can make TAs more similar to the requirement as a bridge through which to find the missing indirect links, the accuracy of trace links may be increased. 

Suppose there is a requirement $RA_1$: \enquote{\emph{All sensitive operations must be authenticated. }}. The task is to identify trace links between this requirement and four use cases. 
\begin{tcolorbox}[
    colback=white,
    colframe=black,
    boxrule=1pt,
    arc=3mm
]
$UC_1$: 

Name: Mobile Verification Code Login

Participants: All users

Main Process:

\begin{enumerate}
    \item The user enters their mobile number to obtain a verification code.
    \item The system sends a 6-digit verification code to the mobile phone.
    \item The user enters the verification code and submits it.
    \item The system verifies the validity of the verification code.
    \item A user session token is created.
\end{enumerate}

\end{tcolorbox}
\begin{tcolorbox}[
    colback=white,
    colframe=black,
    boxrule=1pt,
    arc=3mm
]
$UC_2$: 

Name: Credit Card Payment Processing

Participants: Verified user

Business Value: Complete the settlement of funds for the transaction.

Preconditions:

\begin{enumerate}
    \item The user has a pending payment order in the shopping cart.
    \item The payment gateway service is available.
\end{enumerate}
Main Process:

\begin{enumerate}
    \item The user navigates to the payment page and selects the credit card payment method.
    \item The user enters credit card information (card number, expiration date).
    \item The system encrypts and transmits the payment data to the payment gateway.
    \item Receives and parses the gateway response.
    \item Generate a transaction record (including the gateway transaction ID).
    \item Display the payment result page to the user.
\end{enumerate}
\end{tcolorbox}
\begin{tcolorbox}[
    colback=white,
    colframe=black,
    boxrule=1pt,
    arc=3mm
]
$UC_3$: 

Name: Email and Password Login

Participants: All users

Main Process:

\begin{enumerate}
    \item The user enters their email and password.
    \item The system verifies the password hash.
    \item A user session token is created.
\end{enumerate}
\end{tcolorbox}
\begin{tcolorbox}[
    colback=white,
    colframe=black,
    boxrule=1pt,
    arc=3mm
]
$UC_4$:

Name: Session Activity Validation  

Participants: System security module  

Pre Conditions: Any request for sensitive operations (Such as payment, personal information modification). 

Main Process: 
\begin{enumerate}
    \item Check the last active time of the user's session.
    \item If the time since the last operation is $\leq$ 15 minutes → Allow the operation to continue.
    \item If more than 15 minutes have passed → Trigger UC-AUTH-002 (mandatory re-authentication).
\end{enumerate}
\end{tcolorbox}

\begin{figure}[ht]
    \centering
    \includegraphics[width=1\textwidth]{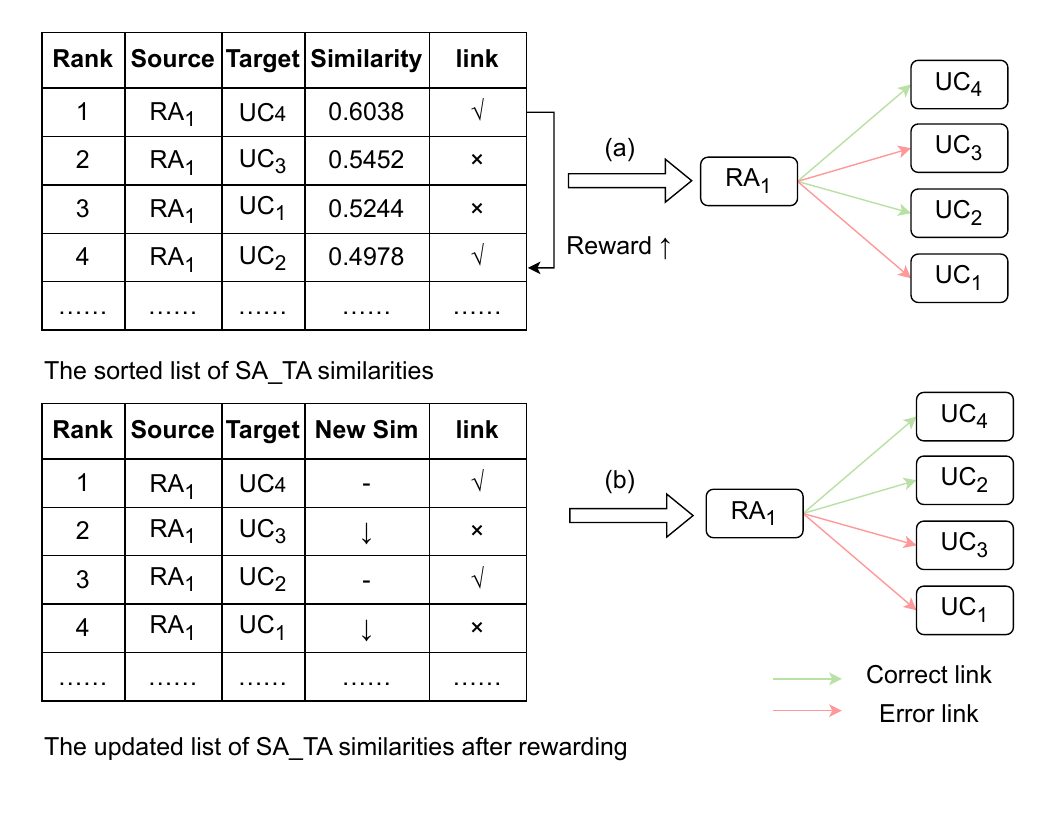}
    \caption{A motivating example}
    \label{fig1}
\end{figure}

Figure~\ref{fig1} shows the proces of identifying trace links between $RA_1$ and the four use cases. 

Although $UC_2$ (Credit Card Payment Processing) exhibits low textual-level semantic similarity scores with $RA_1$ (Security Core Requirement)---sharing only indirect terminological connections such as \enquote{authenticated user}---its implementation logic implicitly invokes $UC_4$ (Session Liveliness Verification) to fulfill $RA_1$ compliance. Specifically, when processing payment requests, $UC_4$ automatically performs session validity checks as an authentication middleware, triggering forced re-authentication upon detecting 15 minutes of inactivity. This functional dependency establishes $UC_4$ as a critical intermediary artifact in the requirement implementation link, thereby justifying the enhanced similarity ranking adjustment for $UC_2$ in section (a) of Figure~\ref{fig1}.

Conversely, while $UC_3$ (Email Password Login) and $UC_1$ (SMS Verification Login) demonstrate higher textual similarity with $RA_1$ through shared core terminology (\enquote{authentication}, \enquote{session tokens}), they fundamentally represent generic authentication mechanisms rather than directly addressing $RA_1$'s specific constraint of \enquote{real-time verification for sensitive operations}. Such foundational use cases should interface with security requirements through system-wide architectural safeguards rather than through explicit trace links. Consequently, we implement appropriate similarity score downweighting for these generic use cases in section (b) of Figure~\ref{fig1} to mitigate misleading associations.

According to this example, T-SimCSE is motivated by three ideas: 
\begin{enumerate}
\item  Correct trace links are more likely to exist in two artifacts (i.e. SA and TA) with high semantic similarity; 
\item  If a trace link exists between a TA and a SA, then the correct trace links to this SA are more likely to exist in other TAs with high semantic similarity with this target artifact;
\item  Compared to a TA similar to many artifacts, a TA that is only similar to only a few artifacts is more likely to have a correct trace link with the SA.
\end{enumerate}

\subsection{Rationale for Selecting SimCSE}\label{subsec3.2}


To support recovery of traceability links between the requirement and NL artifacts, it is essential to employ a model that can generate high-quality semantic embeddings. In this study, we choose the supervised variant of SimCSE~\citet{gao2021simcse} as our similarity computation model. Concretely, we employ the pre-trained princeton-nlp/sup-simcse-roberta-large model from HuggingFace~\footnote{https://huggingface.co/princeton-nlp/sup-simcse-roberta-large}, which has been trained on natural language inference datasets. We do not perform any additional training or fine-tuning on our traceability datasets.

SimCSE leverages contrastive learning to train sentence encoders that bring semantically similar sentence pairs closer in the embedding space while pushing dissimilar pairs farther apart. SimCSE builds upon RoBERTa~\citet{liu2019roberta}, a robustly optimized variant of BERT~\citet{devlin2019bert}, which enhances generalization by using longer training times and larger corpora.

We select SimCSE for this work based on the following advantages:
\begin{enumerate}
\item \textbf{Semantic robustness}: By explicitly modeling both positive and negative pairs, SimCSE effectively distinguishes subtle differences between semantically similar requirements, which is essential for the traceability link recovery task.
\item \textbf{Expression diversity}: Requirements often express the same meaning in various ways. SimCSE brings different forms with the same semantics closer in the embedding space, thus improving the matching accuracy.
\item \textbf{Data efficiency}: Supervised SimCSE requires relatively few labeled data to achieve high performance, which aligns with the limited availability of manually annotated traceability datasets.
\item \textbf{Deployment practicality}: Compared to larger PLMs, SimCSE is more lightweight and easier to deploy in real-world software engineering environments.
\end{enumerate}

In summary, SimCSE offers a well-balanced trade-off between performance, interpretability, and practicality, making it an ideal choice for semantic similarity computation in the context of requirements traceability.

\subsection{Concepts}\label{subsec3.3}

The relevant concepts involved in this paper are as follows:

\textbf{1) Requirements trace link~\citet{niu2012enhancing}:} A requirements trace link is a means used in software engineering and systems engineering to link requirements with their associated outputs (e.g., documentation, code, test cases) from different development phases such as design, development, testing, and maintenance activities. Specific examples are demonstrated in Section~\ref{subsec3.1}.

\textbf{2) Source Artifact (SA)~\citet{lamsweerde2009requirements}:} A source artifact is the originating artifact from which traceability links are established. In this paper,  the source artifact specifically refers to the requirement artifact, e.g. use cases or high-level requirements.

\textbf{3) Target Artifact (TA)~\citet{sundaram2010assessing}:} A target artifact is the dependent artifact highly associated with the source artifact. In this paper, the target artifact specifically refers to the NL artifact, including low-level requirements, design definitions, test cases, etc. 

\textbf{4) High Probability Target Artifacts (HPTA):} Given an ordered similarity list between a SA and TAs, we define a HPTA as the top $k_1$ of TAs that are considered more likely to have trace links with the SA.

\textbf{5) To-be-Rewarded Target Artifacts (TRTA):} For a given HPTA, we first calculate its similarity with all other TAs. The top $k_2$ of TAs, ranked by similarity, are considered to have a high degree of similarity with the HPTA. These artifacts are defined to as TRTAs.

\textbf{6) Specificity:} Specificity denotes the degree of similarity between a TA and other TAs. If a TA has a close relationship with many TAs, it will have a low specificity. Conversely, a TA will have a high specificity if it is closely related to only a few TAs.

\textbf{7)  Final link list:} The final link list refers to an ordered list of TAs associated with the SA by calculating the similarity value and specificity value. Trace links are created between the SA and the top-K TAs in the final link list.

\section{The T-SimCSE approach}\label{sec4}

\begin{figure}[ht]
    \centering
    \includegraphics[width=1\textwidth]{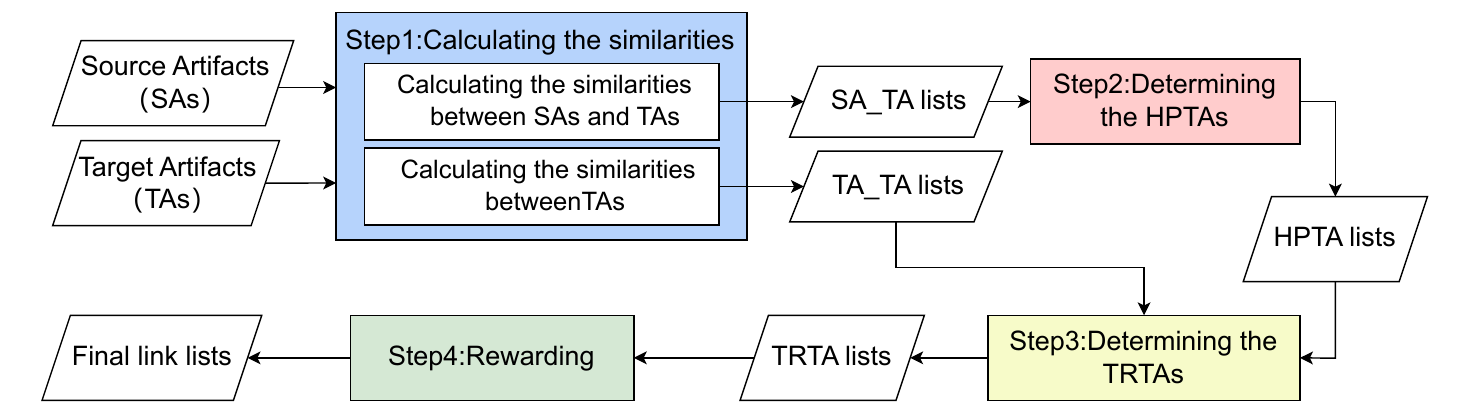}
    \caption{The process of T-SimCSE}
    \label{fig2}
\end{figure}

Motivated by the aforementioned three ideas in Section~\ref{sec3}, the T-SimCSE approach is divided into four main steps, as shown in Figure~\ref{fig2}: 
\begin{enumerate}
    \item Calculating the similarity between the SAs and TAs, and then generating sorted lists of SA\_TA similarities (SA\_TA lists). Meanwhile, we calculate the similarity between TAs and then generate sorted lists of TA\_TA similarities (TA\_TA lists).
    
    \item Determining the HPTAs. HPTAs are more likely to be correctly traceable TAs compared to other TAs. For each SA, a HPTA list is generated in this step. 
    
    \item Determining the TRTAs.  For each HPTA, a TRTA list is generated in this step and those TRTAs will be rewarded in the next step. 
    
    \item Rewarding. In order to differentiate the importance of each TRTA and give them different rewards, we calculate the specificity of each TRTA, which determines its weight in the rewarding phase. If a TRTA has a close relationship with many TAs, it will have low specificity. In contrast, a TRTA will be assigned a high specificity value if it is closely related to only a few TAs. TRTAs that have higher specificity values will receive more rewards during the rewarding phase. A rewarding strategy will be applied to each TRTA for reordering them.  
\end{enumerate}

\subsection{Calculating the similarities}\label{subsec4.1}

First, we use SimCSE to convert the contents of the SAs and TAs into sentence embedding. Next we use the following Formula~\ref{sim_SA} to calculate the similarities (cosine distance) between SAs and TAs, and use Formula~\ref{sim_TA} to calculate the similarities (cosine distance) between TAs.

\begin{equation}
    \text{cosine\_dist} (SA, TA) = 1 - \frac{SA \cdot TA}{\|SA\| \cdot \|TA\|}\label{sim_SA}
\end{equation}

\begin{equation}
    \text{cosine\_dist} (TA, TA) = 1 - \frac{TA \cdot TA}{\|TA\| \cdot \|TA\|}\label{sim_TA}
\end{equation}
where SA is the embedding of one sentence, TA is the embedding of another sentence, and $||SA||$ refers to the length of the embedding.
Suppose a system has a SA set SA\_set = \{$SA_1$, $SA_2$, ..., $SA_n$\} and a TA set TA\_set = \{$TA_1$, $TA_2$, ..., $TA_m$\}. For each SA, we calculate its similarity to the \textit{m} TAs and sort them from highest to lowest according to their similarity values, thus obtaining SA\_TA lists. For each TA, we calculate its similarity with the remaining \textit{m-1} TAs and sort these similarity scores in descending order, ultimately generating TA\_TA lists.

\subsection{Determining the HPTAs}\label{subsec4.2}

For each SA, there is a sorted SA\_TA list in descending order of similarity. The top $k_1$ TAs in this list are selected as HPTAs, which are more likely to be correctly traceable TAs. These HPTAs are used to select TRTAs in Step 3. The selection of threshold $k_1$ will be discussed in Section~\ref{sec6}. 

\subsection{Determining the TRTAs}\label{subsec4.3}

In T-SimCSE, it is crucial to determine which TAs are qualified for rewards, i.e. determining the TRTAs. Specifically, for a given HPTA, we need to first calculate the similarity between the HPTA and all other TAs. The top $k_2$ TAs have a high degree of similarity with its corresponding HPTA and thus have a high degree of similarity with the SA. We treat these top $k_2$ TAs as TRTAs and reward them in Step 4. To ensure adaptability across different domains and datasets with varying characteristics, we employ a dynamic threshold $k_2$ to determine the TRTAs, which will be also discussed in Section~\ref{sec6}.

\subsection{Rewarding}\label{subsec4.4}

1) Calculating the specificity of each TRTA

\begin{figure}[ht]
    \centering
    \includegraphics[width=0.85\textwidth]{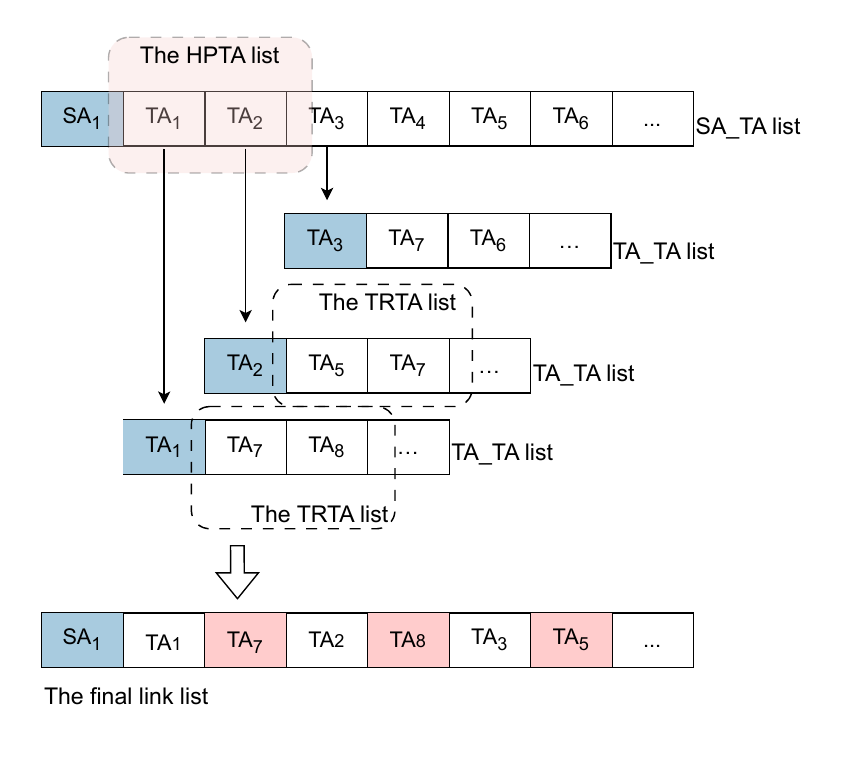}
    \caption{A rewarding example}
    \label{fig_reward_ex}
\end{figure}

As shown in Figure~\ref{fig_reward_ex}, after we determine the HPTA lists according to the value $k_1$, for each HPTA, we calculate the similarity between the HPTA and other TAs, and use $k_2$ to obtain a set of TRTA lists.  

Each of these TRTAs corresponds to a \textit{count} value , denoted as $count_i$. 

We use \textit{count} to denote the number of times a given TRTA appears in the top $k_2$ rankings of TA\_TA lists. 

Taking the example in Figure~\ref{fig_reward_ex}, suppose $k_2$ results in selecting the top 2 entries from each list, then $TA_7$ is selected as a TRTA. Since $TA_7$ appears three times for $TA_1$, $TA_2$, and $TA_3$, the \textit{count} for $TA_7$ is 3. $TA_8$ and $TA_5$ appear only once in the top 2 of $TA_1$ and $TA_2$'s lists, respectively, so their \textit{counts} are 1. If $TA_5$, $TA_7$, or $TA_8$ also appear in the top 2 of $TA_4$’s TA\_TA list, their corresponding \textit{counts} will be incremented accordingly.
Based on the \textit{count} value, we can calculate the specificity of each TA, which is calculated as follows:

\begin{equation}
    Spe_i = log\frac{m-1}{count_i}\label{spe}
\end{equation}

Where $Spe_i$ denotes the specificity of the $i$th TRTA,

$count_i$ denotes the number of times the $i$th TRTA appears in all top $k_2$ TA\_TA lists and \textit{m} denotes the total number of TAs. Since a TRTA can be identified as top $k_2$ at most \textit{m-1} times, we set the numerator to \textit{m-1}. Formula~\ref{spe} shows that if a TRTA has a close relationship with many TAs, it will have a high value of \textit{count} as well as a low specificity.

2) Rewarding TRTAs according to their specificity values

T-SimCSE applies the obtained specificity values to reward the TRTAs. The rewarding value is calculated as below. Assuming that a HPTA has \textit{r} TRTAs, we re-calculate the similarity of the \textit{i}th TRTA by using the following formula:

\begin{equation}
    Reward_i = (Sim_{first} - Sim_{origin}) * (\frac{Spec_i}{Spec_1+Spec_2+...+Spec_r})\label{rewarding}
\end{equation}

\begin{equation}
    Sim_{new} = Sim_{origin} + Reward_i \label{newSim}
\end{equation}

Where $Reward_i$ denotes the rewarding value should be given to the \textit{i}th TRTA, $Sim_{origin}$ indicates the similarity value of this TRTA in the SA\_TA list, $Sim_{first}$ indicates the similarity value of the first ranked TA in the SA\_TA list. $Sim_{new}$ indicates the similarity of this TRTA after applying the reward. For example, in the case shown in Figure ~\ref{fig_reward_ex}, the first ranked TA corresponding to all TRTAs is $TA_1$ as $TA_1$ is most similar to the SA in the SA\_TA list. $Spe_i$ indicates the specificity of the \textit{i}th TRTA. 

Formula~\ref{newSim} is designed based on the assumption that a TA with a high specificity is more likely to create a valid trace link with the SA than a less specific TA when both of them are considered to be semantically close artifacts. 

Finally, we reorder the SA\_TA list by applying the new similarity values to TRTAs as rewards, and then sort all TAs in descending order of their similarity values. For example, as shown in Figure~\ref{fig_reward_ex}, $TA_1$ and $TA_2$ serve as HPTAs of $SA_1$, which means that the TRTAs corresponding to $TA_1$ and $TA_2$ are eligible for rewards. Based on the previous step, we obtain the specificity values for $TA_5$, $TA_7$, and $TA_8$. Then, the updated similarity scores are calculated using Formula~\ref{rewarding} and Formula~\ref{newSim}. By reordering the similarity values of all TAs in descending order, we obtain the final link list. In the final link list, the TAs highlighted in red represent the TRTAs that have been rewarded.

\section{Experiment setup}\label{sec5}

In this section, in order to explore the validity of the T-SimCSE approach, we propose the following three research questions for discussion:

RQ1: For the two thresholds $k_1$ and $k_2$, what values are optimal in T-SimCSE?

RQ2: How effective is the T-SimCSE approach compared to the baselines?

RQ3: How effective is the T-SimCSE approach compared to the prior approaches in requirement traceability?

RQ4: Is the rewarding strategy for the TAs effective?

The experimental code and data have been made public~\footnote{https://github.com/AltmanCat/T-SimCSE.git} for reproduction and research purposes.

\subsection{Datasets}\label{subsec5.1}

We evaluated T-SimCSE on ten datasets. The EasyClinic, GANNT, IceBreaker, CCHIT, EBT, InfusionPump, CM1 and WARC datasets are sourced from the official CoEST platform\footnote{http://sarec.nd.edu/coest/datasets.html}, a benchmark repository widely adopted in requirement traceability recovery studies. Details of the datasets are shown in Table~\ref{tab:datasets}.

\begin{table}[ht]
\centering
\caption{Datasets Information Summary}
\normalsize
\label{tab:datasets}
\setlength{\tabcolsep}{1.5pt}
\begin{tabular}{lcccc}
\toprule
\textbf{Dataset} & \textbf{\shortstack{Source \\ artifact (No.)}} & \textbf{\shortstack{Target \\ artifact (No.)}} & \textbf{Links} & \textbf{Domain} \\
\midrule
EasyClinic      & Use Cases (30)                & Test Cases (63)               & 251            & Healthcare                     \\
GANNT           & HR\tnote{1}  (17)        & LR\tnote{2}  (69)         & 68             & Project Management             \\
    CM1          & HR (22)              & LR (53)              & 230  & Aerospace  \\
 CCHIT        & Req (116)    & RC\tnote{6} (1064) & 587  & Healthcare   \\
    MODIS        & HR (19)              & LR (49)              & 41 &  Aerospace \\
Dronology       & Req\tnote{4}  (99)             & DD\tnote{3}  (211)      & 220            & Aerospace                       \\
    WARC         & HR (63)              & LR (89)              & 136 &	Web Tools  \\
    EBT          & Req (40)     & Test cases (25)       & 51  & Transportation  \\
    InfusionPump & Req (126)    & Components (21)       & 131  & Healthcare  \\
IceBreaker      & CD\tnote{5}  (73)           & Req\tnote{4}  (201)            & 457            & Public Works             \\
\bottomrule
\end{tabular}
\begin{tablenotes}
    \small
    \item[~] \textsuperscript{1}High-level requirements;
    \textsuperscript{2}Low-level requirements;
    \textsuperscript{3}Design definitions;
    \textsuperscript{4}Requirements;
    \textsuperscript{5}Class diagrams descriptions;
    \textsuperscript{6}Regulatory codes;
    
\end{tablenotes}
\end{table}

\subsection{Baselines}\label{subsec5.2}

To comprehensively assess the performance of our approach, we employ five baselines: the original SimCSE approach and four baseline approaches — a VSM-based approach, a LSI-based approach, a BERT-based approach, and a Word2Vec-based approach. In our experimental setup, these four baselines replace SimCSE with different embedding strategies, enabling a thorough evaluation across diverse embedding techniques.

1) As a baseline approach, SimCSE establishes trace links by directly computing the semantic similarity between SAs and TAs without incorporating any reward mechanism. In contrast, the proposed T-SimCSE enhances SimCSE by integrating a rewarding strategy. By comparing the performance of both approaches, the impact of the reward component on link quality can be intuitively verified, providing empirical support for the rationale of the proposed strategy.

2) The VSM-based approach~\citet{antoniol2002recovering} has been widely applied in the domain of requirements traceability. Its core principle involves mapping code and documentation into vector representations through vocabulary construction, enabling developers to efficiently retrieve relevant information, track code evolution, and comprehend system structures. This study extends the VSM model by first generating artifact vectors using TF-IDF weighting and applying cosine similarity for preliminary link filtering. Subsequently, a rewarding strategy is introduced to refine similarity weights, ultimately selecting the high-confidence links to form the trace link.

3) The LSI-based approach~\citet{marcus2003recovering} aims to capture latent semantic structures in textual data through mathematical modeling. By employing linear algebra techniques such as Singular Value Decomposition (SVD), high-dimensional term-document matrices are mapped into a lower-dimensional latent semantic space, reducing data sparsity and uncovering implicit relationships between terms and documents. The implementation of LSI in this study follows these steps: first, SVD is applied to reduce the dimensionality of the term matrix of artifacts, obtaining latent semantic vector representations; then, sorted link lists are generated using cosine similarity combined with a rewarding strategy; finally, links with higher similarity scores are retained as the trace link.

4) BERT-based approaches have been extensively explored in requirements traceability research. For example, T-BERT~\citet{lin2021traceability} constructs software trace links using a pre-trained BERT model, which learns word distributions from large-scale unlabeled natural language and programming language documents. Moreover, by fine-tuning on real-world open-source projects, T-BERT has been applied to recovering trace links between issues and commits. In this study, the Sentence-BERT model (sentence-transformers/bert-base-nli-mean-tokens\footnote{https://huggingface.co/sentence-transformers/bert-base-nli-mean-tokens}~\citet{reimers2019sentencebert}) is employed to encode artifact content into dense vectors. Semantic similarity is calculated using cosine similarity, and a rewarding strategy is incorporated to dynamically adjust the similarity weights, ultimately generating the top trace links.

5) The Word2Vec-based approach first maps words in software artifacts into real-valued vectors to represent words as continuous vectors and capture their semantic relationships. For instance, Fine-Grained Traceability Link Recovery (FTLR)~\citet{hey2021improving} considers sentences in requirement documents and methods in source code as the smallest representation units. These elements are transformed into vector forms via word embeddings, and trace links are recovered through similarity computations. In this study, the pre-trained GoogleNews-vectors-negative300\footnote{https://huggingface.co/LoganKilpatrick/GoogleNews-vectors-negative300} model is utilized to embed artifact content. To obtain document-level representations, we adopt a mean pooling~\citet{arora2017simple} strategy that averages the word embeddings of non-stopword tokens in each artifact. A rewarding strategy is then applied to refine similarity weights, ultimately selecting the top links to construct the trace link.

\subsection{Prior approaches}\label{subsec5.3}

In recent years, numerous approaches have been proposed to improve the effectiveness of requirements traceability by addressing challenges such as vocabulary mismatch, semantic representation, and retrieval accuracy. The following representative approaches illustrate different technical directions, ranging from embedding-based models and query expansion to graph mining and LLM integration. We briefly summarize these approaches below for comparison.

1) S2Trace ~\citet{chen2019enhancing}: S2Trace is an unsupervised requirements traceability approach to enhance the generation of trace links by leveraging sequential semantic information embedded in software artifacts. It begins by automatically extracting distance-aware sequential patterns from artifact texts. These patterns are then combined with original textual terms to learn document embeddings using the Doc2Vec model. To further improve the efficiency of feature representation, S2Trace applies Principal Component Analysis (PCA) to reduce redundancy in the high-dimensional embedding space, producing a compact set of principal components. Finally, it computes similarity scores between requirement and code artifact embeddings in this reduced space, ranks the candidate pairs, and selects the top-K links as the final traceability results.

2) WQI ~\citet{zhao2017improved}: The WQI approach captures semantic associations between terms using a word embedding model and employs query expansion techniques to mitigate the vocabulary gap common in traditional information retrieval approaches. Additionally, it incorporates an IDF-based strategy to emphasize the importance of key terms in the dataset, resulting in more accurate rankings of trace links.

3) LiSSA-CoT-GPT4o and LiSSA-CoT-Llama3.1 ~\citet{hey2025requirements}: Since the original paper did not assign a specific name to the approach, we refer to the best-performing approach presented in the paper as LiSSA-CoT-GPT4o and LiSSA-CoT-Llama3.1 for clarity. It adopts a Retrieval-Augmented Generation (RAG) framework for requirements traceability, first retrieving semantically relevant target requirements using embedding-based retrieval. Then, it leverages the LLM GPT-4o with Chain-of-Thought prompting to generate traceability decisions.

4) GeT2Trace ~\citet{chen2021}: GeT2Trace is an enhanced requirements traceability recovery approach based on graph mining and representation expansion. Its core idea is to improve the accuracy of semantic representations by exploiting the latent word co-occurrence and word order information embedded in software artifacts. Specifically, the approach first employs a Graph-of-Word model to transform software artifacts into textual networks. It then mines co-occurring terms and frequent subgraphs at both the project-level and artifact-level granularity, which respectively address the issues of terminology mismatch and word order loss. Based on this mined information, GeT2Trace expands the textual features of the artifacts and learns enhanced document embeddings using the Doc2Vec model. Finally, trace links between requirements are determined by computing the cosine similarity between the corresponding artifact vectors.

Since none of the aforementioned approaches provide reproducible code packages, we conduct comparative analysis based on the data reported in the respective publications.

\subsection{Metrics}\label{subsec5.4}

The metric of precision, recall, AP, MAP and $F_\beta$-score are used to evaluate T-SimCSE. Below are their calculation formulas.

\begin{equation}
Precision  = \frac{|correct \cap retrieved|}{|retrieved|}
\end{equation}

\begin{equation}
Recall  = \frac{|correct \cap retrieved|}{|correct|}
\end{equation}

where \textit{correct} indicates the number of correct links and \textit{retrieved} is the number of all links retrieved by the traceability recovery approach. A common way of evaluating traceability recovery approaches is to plot Precision-Recall (P-R) curves to compare the magnitude of accuracy obtained at different recall levels. Specifically, the curve is plotted with recall rate as the x-axis and precision as the y-axis. The corresponding precision at different recalls is calculated, and these points are connected to form the curve. 

AP measures the extent to which all relevant TAs for a SA are ranked at the top of the retrieved links, which is calculated as:

\begin{equation}
    AP = \frac{\sum_{r=1}^NPrecision(r)\cdot isRelevant(r)}{|RelevantArtifacts|}
\end{equation}

where \textit{r} is the rank of the TA in the sorted link list, \textit{N} is the number of TAs, \textit{Precision (r)} is the precision rate up to the current ranking, and \textit{isRelevant (r)} is a binary function that takes the value of 1 if the current TA is relevant and 0 otherwise. 

MAP is the average of the APs in a set of SAs and it is calculated as follows:

\begin{equation}
    MAP = \frac{\sum_{q=1}^Q AP}{Q}
\end{equation}

Where \textit{q} is the AP for each individual SA, i.e. the \textit{q}th SA, and \textit{Q} is the total number of SAs.
\begin{equation}
F_\beta = (1 + \beta^2) \times \frac{\text{Precision} \times \text{Recall}}{(\beta^2 \times \text{Precision}) + \text{Recall}}
\end{equation}

In this paper, we chose $\beta = 1$ and $\beta = 2$, corresponding to the $F_1$-score and $F_2$-score, respectively.

\section{Results and Analysis}\label{sec6}

\subsection{\texorpdfstring{RQ1: For the two thresholds $k_1$ and $k_2$, what values are optimal in T-SimCSE?}{RQ1: For the two thresholds k1 and k2, what values are optimal in T-SimCSE?}}\label{subsec6.1}

In Section~\ref{sec4}, we apply the threshold $k_1$ to determine the HPTAs and threshold $k_2$ to determine the TRTAs. In this section, we discuss the selection of parameters $k_1$ and $k_2$ and their impact on model performance.

\begin{figure}[ht]
\centering

\begin{subfigure}[b]{0.20\textwidth}
    \centering
    \includegraphics[width=\linewidth]{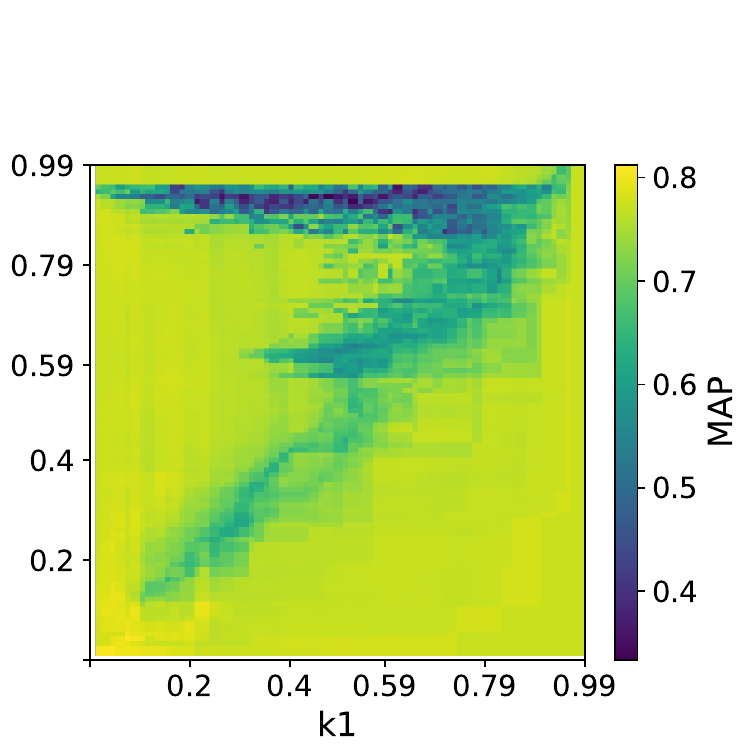}
    \caption{EasyClinic}
    \label{fig4:1}
\end{subfigure}\hfill
\begin{subfigure}[b]{0.20\textwidth}
    \centering
    \includegraphics[width=\linewidth]{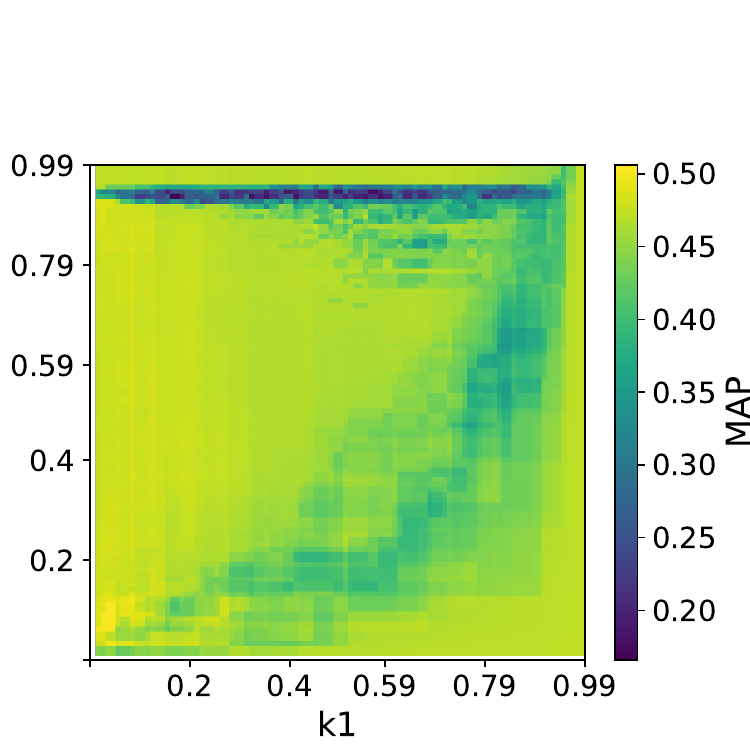}
    \caption{GANNT}
    \label{fig4:2}
\end{subfigure}\hfill
\begin{subfigure}[b]{0.20\textwidth}
    \centering
    \includegraphics[width=\linewidth]{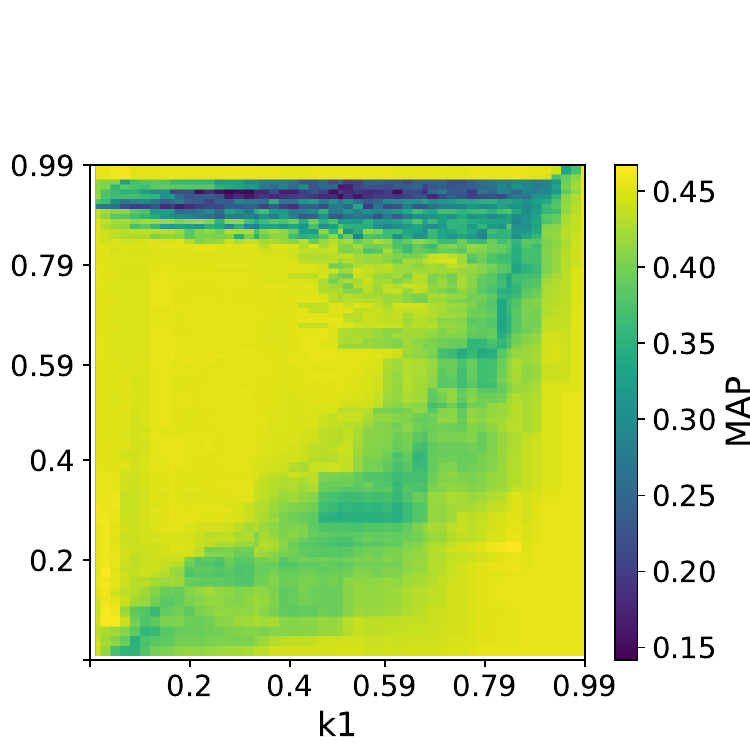}
    \caption{CM1}
    \label{fig4:3}
\end{subfigure}\hfill
\begin{subfigure}[b]{0.20\textwidth}
    \centering
    \includegraphics[width=\linewidth]{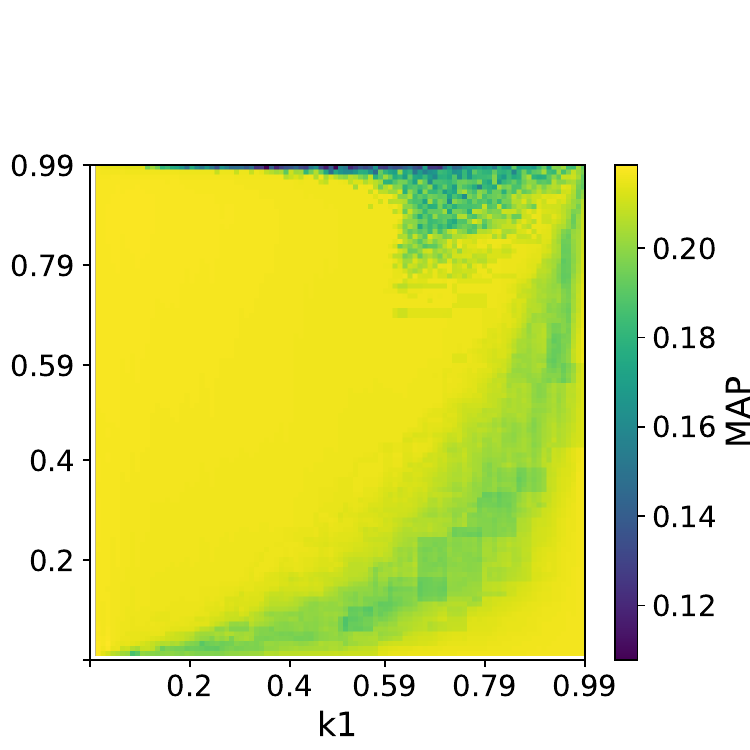}
    \caption{CCHIT}
    \label{fig4:4}
\end{subfigure}\hfill
\begin{subfigure}[b]{0.20\textwidth}
    \centering
    \includegraphics[width=\linewidth]{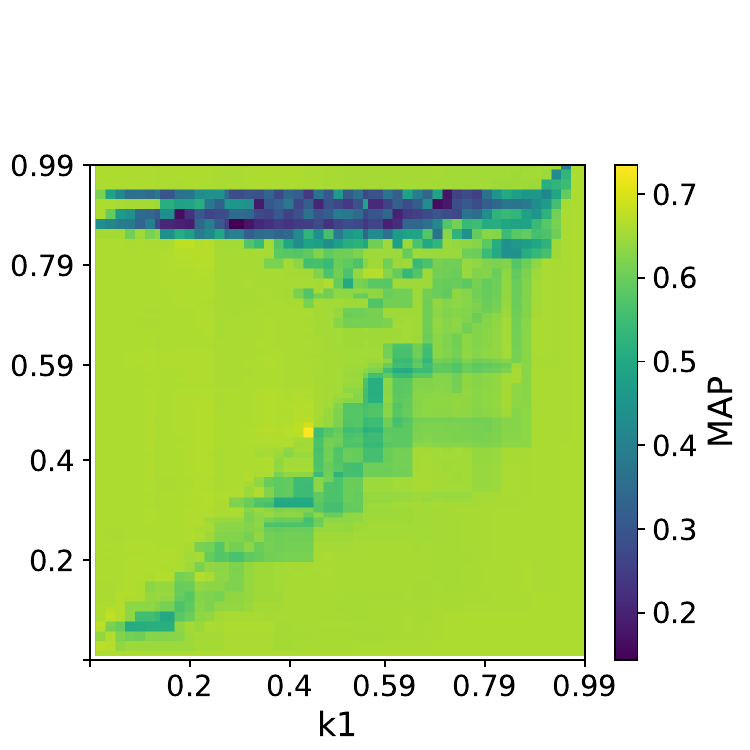}
    \caption{MODIS}
    \label{fig4:5}
\end{subfigure}

\vspace{0.1em}

\begin{subfigure}[b]{0.20\textwidth}
    \centering
    \includegraphics[width=\linewidth]{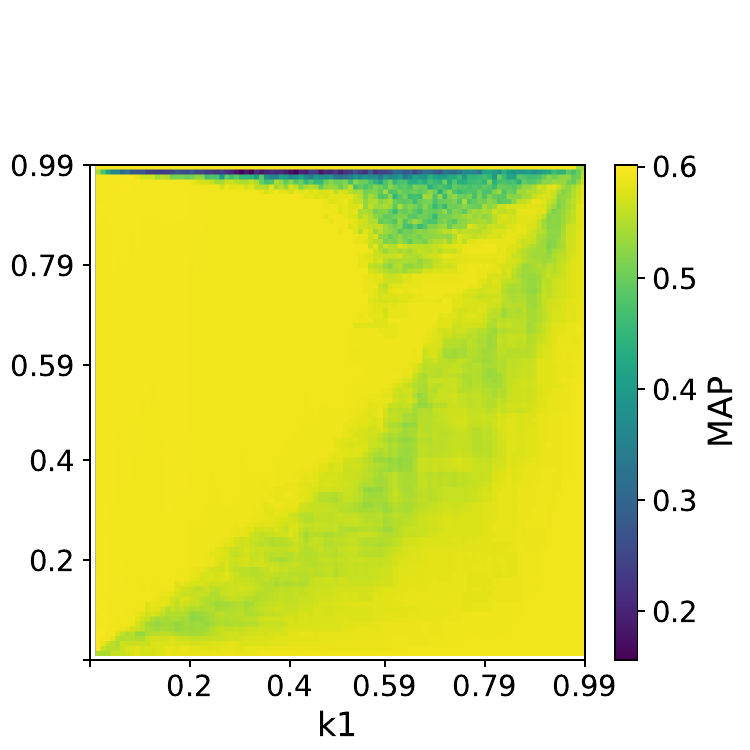}
    \caption{Dronology}
    \label{fig4:6}
\end{subfigure}\hfill
\begin{subfigure}[b]{0.20\textwidth}
    \centering
    \includegraphics[width=\linewidth]{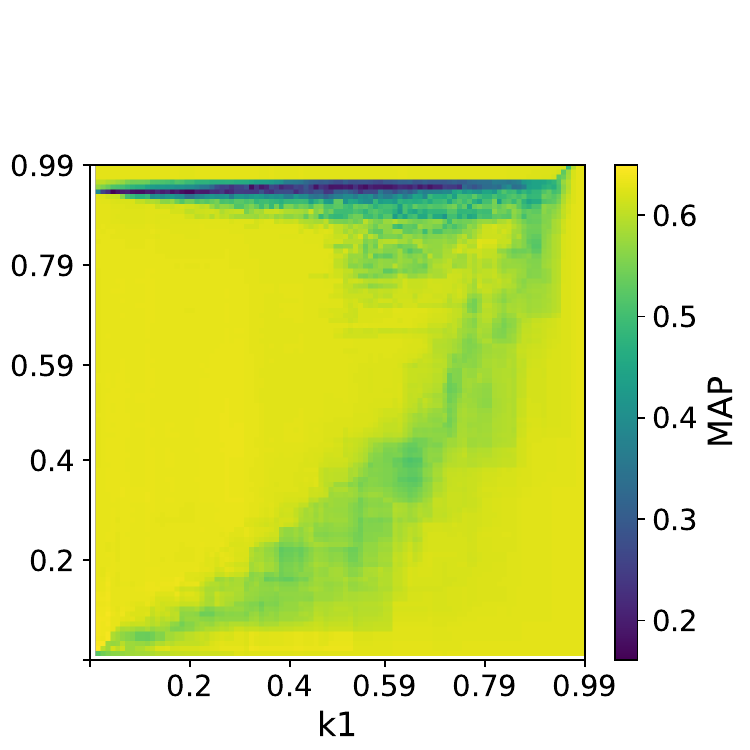}
    \caption{WARC}
    \label{fig4:7}
\end{subfigure}\hfill
\begin{subfigure}[b]{0.20\textwidth}
    \centering
    \includegraphics[width=\linewidth]{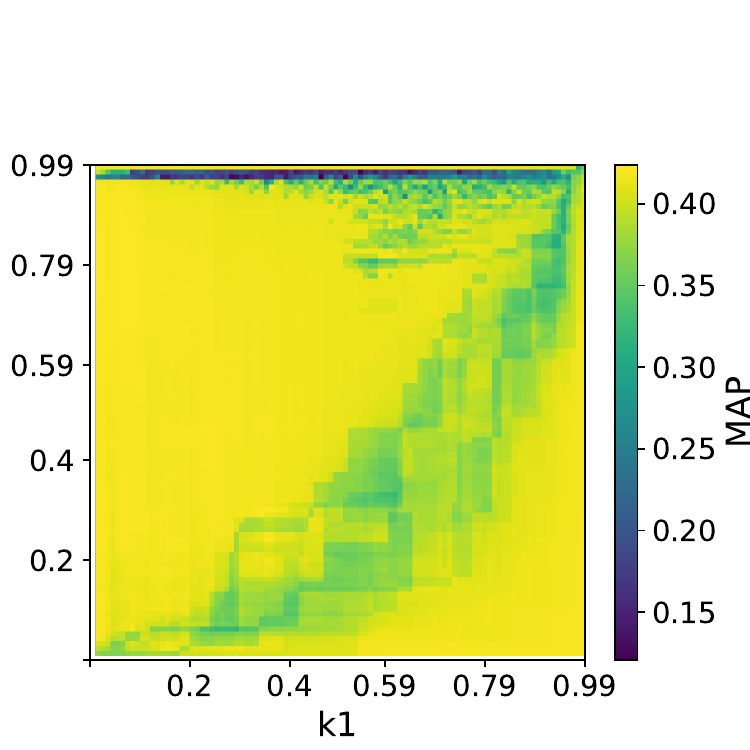}
    \caption{InfusionPump}
    \label{fig4:8}
\end{subfigure}\hfill
\begin{subfigure}[b]{0.20\textwidth}
    \centering
    \includegraphics[width=\linewidth]{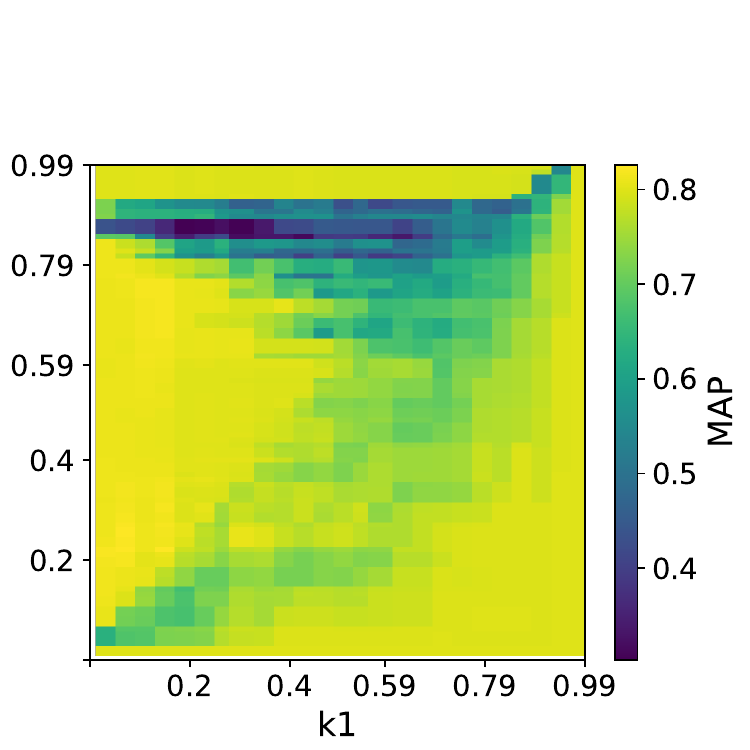}
    \caption{EBT}
    \label{fig4:9}
\end{subfigure}\hfill
\begin{subfigure}[b]{0.20\textwidth}
    \centering
    \includegraphics[width=\linewidth]{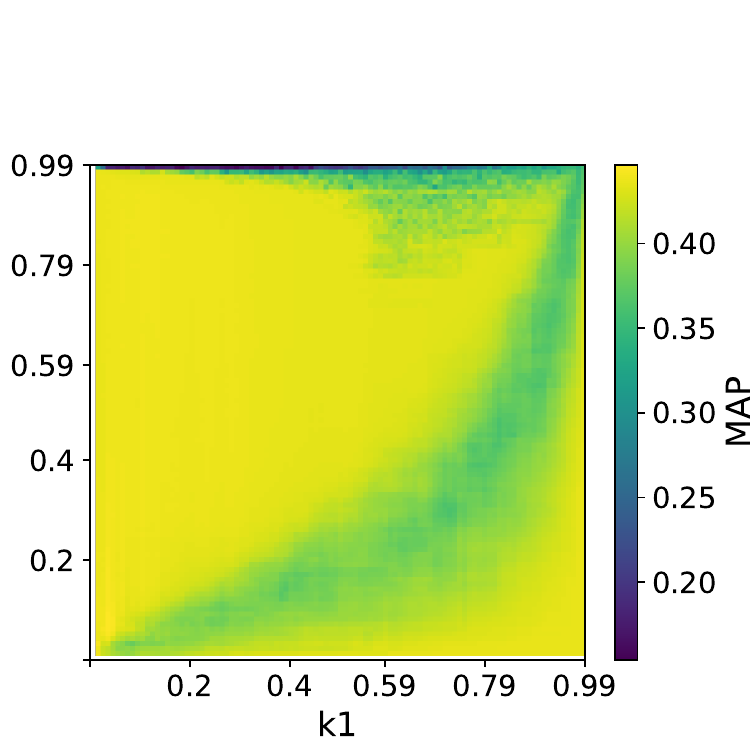}
    \caption{IceBreaker}
    \label{fig4:10}
\end{subfigure}

\caption{Heatmaps of MAP for $k_1$ and $k_2$ in T-SimCSE across ten datasets}
\label{fig-Heatmap}
\end{figure}

For the selection of the threshold values $k_1$ and $k_2$ in T-SimCSE, we adopt a dynamic threshold selection strategy based on grid search to enhance the approach’s performance. Specifically, for the two key thresholds  $k_1$ and $k_2$, we construct a candidate threshold space within the closed interval (0, 1] with 0.01 increments. We then systematically explore all possible threshold combinations through an exhaustive search. For each pair ( $k_1$, $k_2$), we calculate the corresponding MAP score. The optimal threshold configuration is determined as the one that yields the highest MAP value, ensuring a quantitative basis for threshold selection through a full-space search strategy. The results indicate that the optimal values are concentrated within the range (0, 0.2). In future research, we plan to conduct a more fine-grained exploration within this range.

We conducted experiments with T-SimCSE on ten datasets, and Figure~\ref{fig-Heatmap} presents the heatmaps. The results show that the optimal range for $k_1$ is (0, 0.15), while the optimal range for $k_2$ is (0, 0.2). We ultimately selected $k_1 = 0.03$ and $k_2 = 0.08$ as the final configuration.

As observed in the heatmap in Figure~\ref{fig-Heatmap}, the MAP value reaches its maximum in the lower-left corner and gradually decreases as either $k_1$ or $k_2$ increases. This suggests that larger values of $k_1$ or $k_2$ may lead to performance degradation. For certain datasets (e.g., MODIS and GANNT), the MAP value exhibits a more gradual decline, indicating lower sensitivity to threshold variations. In contrast, for other datasets (e.g., IceBreaker and Dronology), the MAP values exhibit steeper gradients, revealing higher sensitivity to threshold selection.

\subsection{RQ2: How effective is the T-SimCSE approach compared to the baselines?}\label{subsec6.2}

To evaluate the effectiveness of the T-SimCSE approach, we compared it with four baseline approaches across ten datasets. We report results in Table~\ref{tab:model_performance}.

Across the ten datasets, T-SimCSE achieved the highest MAP values on eight datasets. The only exceptions were InfusionPump, where Word2Vec slightly outperformed T-SimCSE (0.43 vs. 0.42), and CM1, where LSI (0.70) and VSM (0.62) surpassed T-SimCSE (0.47).

\begin{table}[htbp]
\centering
\caption{The number of calculated MAP, $p$, and Cliff's~$\delta$ evaluating each approach}
\label{tab:model_performance}
\small
\setlength{\tabcolsep}{1.6pt}
\renewcommand{\arraystretch}{1.02}

\begin{tabular}{l@{} *{12}{c}@{}}
\toprule

\multirow{2}{*}{Approach}
& \multicolumn{3}{c}{EasyClinic}
& \multicolumn{3}{c}{GANNT}
& \multicolumn{3}{c}{CM1}
& \multicolumn{3}{c}{} \\ 
\cmidrule(lr){2-4}\cmidrule(lr){5-7}\cmidrule(lr){8-10}
& MAP & $p$ & $\delta$
& MAP & $p$ & $\delta$
& MAP & $p$ & $\delta$
& \multicolumn{3}{c}{} \\
\midrule
T\,-SimCSE
& \textbf{0.81} & -- & --
& \textbf{0.51} & -- & --
& 0.47 & -- & --
& \multicolumn{3}{c}{} \\
Word2Vec
& 0.74 & $<0.01$ & 0.80
& 0.32 & $<0.01$ & 1
& 0.44 & $<0.01$ & 0.60
& \multicolumn{3}{c}{} \\
BERT
& 0.53 & $<0.01$ & 1
& 0.32 & $<0.01$ & 1
& 0.31 & $<0.01$ & 1
& \multicolumn{3}{c}{} \\
LSI
& 0.75 & 0.02 & 0.60
& 0.46 & 0.06 & 0.40
& \textbf{0.70} & $<0.01$ & $-1$
& \multicolumn{3}{c}{} \\
VSM
& 0.72 & $<0.01$ & 1
& 0.39 & $<0.01$ & 1
& 0.62 & $<0.01$ & $-1$
& \multicolumn{3}{c}{} \\

\addlinespace[2pt]
\cmidrule(lr){2-13}
\addlinespace[2pt]

\multicolumn{1}{l}{}%
& \multicolumn{3}{c}{CCHIT}
& \multicolumn{3}{c}{MODIS}
& \multicolumn{3}{c}{Dronology}
& \multicolumn{3}{c}{} \\
\cmidrule(lr){2-4}\cmidrule(lr){5-7}\cmidrule(lr){8-10}
\multicolumn{1}{l}{}%
& MAP & $p$ & $\delta$
& MAP & $p$ & $\delta$
& MAP & $p$ & $\delta$
& \multicolumn{3}{c}{} \\
\midrule
T\,-SimCSE
& \textbf{0.22} & -- & --
& \textbf{0.74} & -- & --
& \textbf{0.60} & -- & --
& \multicolumn{3}{c}{} \\
Word2Vec
& 0.17 & $<0.01$ & 1
& 0.62 & $<0.01$ & 1
& 0.47 & $<0.01$ & 1
& \multicolumn{3}{c}{} \\
BERT
& 0.08 & $<0.01$ & 1
& 0.54 & $<0.01$ & 1
& 0.41 & $<0.01$ & 1
& \multicolumn{3}{c}{} \\
LSI
& 0.19 & 0.19 & 0.20
& 0.56 & $<0.01$ & 1
& 0.57 & $<0.01$ & 1
& \multicolumn{3}{c}{} \\
VSM
& 0.19 & $<0.01$ & 0.80
& 0.57 & $<0.01$ & 1
& 0.54 & $<0.01$ & 1
& \multicolumn{3}{c}{} \\

\addlinespace[2pt]
\cmidrule(lr){2-13}
\addlinespace[2pt]

\multicolumn{1}{l}{}%
& \multicolumn{3}{c}{WARC}
& \multicolumn{3}{c}{InfusionPump}
& \multicolumn{3}{c}{EBT}
& \multicolumn{3}{c}{IceBreaker} \\
\cmidrule(lr){2-4}\cmidrule(lr){5-7}\cmidrule(lr){8-10}\cmidrule(lr){11-13}
\multicolumn{1}{l}{}%
& MAP & $p$ & $\delta$
& MAP & $p$ & $\delta$
& MAP & $p$ & $\delta$
& MAP & $p$ & $\delta$ \\
\midrule
T\,-SimCSE
& \textbf{0.65} & -- & --
& 0.42 & -- & --
& \textbf{0.83} & -- & --
& \textbf{0.45} & -- & -- \\
Word2Vec
& 0.48 & $<0.01$ & 1
& \textbf{0.43} & 0.38 & 0.00
& 0.60 & $<0.01$ & 1
& 0.41 & $<0.01$ & 1 \\
BERT
& 0.49 & $<0.01$ & 1
& 0.27 & $<0.01$ & 1
& 0.64 & $<0.01$ & 1
& 0.26 & $<0.01$ & 1 \\
LSI
& 0.62 & $<0.01$ & 1
& 0.41 & 0.92 & 0.00
& 0.68 & $<0.01$ & 1
& 0.40 & $<0.01$ & 1 \\
VSM
& 0.56 & $<0.01$ & 1
& 0.41 & 1.00 & $-0.40$
& 0.62 & $<0.01$ & 1
& 0.39 & $<0.01$ & 1 \\
\bottomrule
\end{tabular}
\end{table}

To further assess the robustness of these improvements, we conducted Wilcoxon signed-rank tests with Cliff’s~$\delta$ effect size based on ten precision values sampled at recall levels from 0.1 to 1.0 (step size = 0.1). The magnitude of Cliff’s~$\delta$ is interpreted following the guidelines proposed by Romano et al.~\citet{romano2006appropriate}. The results are summarized as follows:

\begin{itemize}
    \item \textbf{Compared with BERT}: T-SimCSE achieved statistically significant improvements on all 10 datasets ($p < 0.05$), with \textit{large} effect sizes ($\delta > 0$) throughout, demonstrating consistent superiority.
    
    \item \textbf{Compared with Word2Vec}: T-SimCSE was significantly better on 9 out of 10 datasets, with all significant differences being \textit{large} ($\delta > 0$). The only exception was \textit{InfusionPump}, where the difference was not significant ($\delta \approx 0$, negligible).
    
    \item \textbf{Compared with LSI}: T-SimCSE achieved significant improvements on 6 datasets, all with \textit{large} effect sizes. On GANNT ($p = 0.064$, medium), CCHIT (small), and InfusionPump (negligible), the differences were not significant. Notably, on CM1 the effect size $\delta < 0$, indicating that LSI was significantly better than T-SimCSE.
    
    \item \textbf{Compared with VSM}: T-SimCSE significantly outperformed VSM on 8 datasets, all with \textit{large} effect sizes ($\delta > 0$). On \textit{InfusionPump}, the difference was not significant, with $\delta < 0$ suggesting VSM had a slight but non-significant advantage. On \textit{CM1}, however, VSM was significantly better than T-SimCSE ($\delta < 0$, large).
\end{itemize}

To further examine the differences among the approaches in terms of MAP values, we conducted the Friedman omnibus test, followed by the Nemenyi post-hoc test~\citet{DellAydeDalp2023}. The Friedman omnibus test ($\chi^{2}=25.12$, $p<0.001$) revealed statistically significant differences among the approaches. The Nemenyi post-hoc test showed that T-SimCSE significantly outperformed Word2Vec ($p=0.016$), BERT ($p<0.001$), and VSM ($p=0.038$), whereas no significant difference was observed between T-SimCSE and LSI ($p=0.618$). According to the average ranks, T-SimCSE achieved the best overall performance ($1.3$), followed by LSI ($2.3$), while BERT ranked the lowest ($4.6$). 

From the above analysis, it can be concluded that the experimental results of T-SimCSE are significantly superior to those of the BERT-based approach. This is primarily due to the following reasons:

\begin{enumerate}
    \item While Sentence-BERT optimizes sentence representations through supervised learning (e.g., NLI tasks), bringing embeddings of similar sentences closer and partially alleviating BERT's original anisotropy~\citet{gao2019representation, wang2019improving} issue, residual anisotropy may still persist in its vector space when processing complex semantics or out-of-domain data.
    \item Although Sentence-BERT mitigates the impact of word frequency on overall sentence representation through sentence-level training objectives, it may still underperform when low-frequency words critical to semantics (e.g., technical terms) in input texts lack sufficient training exposure in Sentence-BERT.
\end{enumerate}

In summary, T-SimCSE demonstrates consistent and statistically significant advantages over BERT and Word2Vec across almost all datasets, while also outperforming LSI and VSM on the majority of cases.

\subsection{RQ3: How effective is the T-SimCSE approach compared to the prior approaches in requirement traceability?}\label{subsec6.3}

We conducted comparative experiments on seven datasets to systematically evaluate the performance of the proposed T-SimCSE approach against several prior approaches, including LiSSA-CoT-GPT4o, LiSSA-CoT-Llama3.1, WQI, S2Trace, and GeT2Trace. For the remaining three datasets (InfusionPump, EBT, and IceBreaker), we exclude them from the main analysis of RQ3 because prior approaches did not provide published results and no reproducible implementations are available.  We use the $k_1$ and $k_2$ selected in RQ1 and select the top six in the final link list as the trace links. The results are summarized in Table~\ref{tab4}.

\begin{table}[htbp]
    \centering
    \caption{Performance metrics across seven datasets}
     \label{tab4}
    \begin{tabular}{l l c c c c}
    \toprule
      Dataset          & Method                    & Precision      & Recall         & $F_1$             & $F_2$            \\
    \midrule
    \multirow{4}{*}{EasyClinic} 
        & T-SimCSE     & \textbf{0.76} & 0.60 & \textbf{0.67} & 0.63 \\
        & WQI          & 0.50 & \textbf{0.87} & 0.63 & \textbf{0.76} \\
        & S2Trace      & 0.47 & 0.74 & 0.58 & 0.66 \\
        & GeT2Trace    & 0.48 & 0.77 & 0.59 & 0.69 \\
    \midrule
    \multirow{6}{*}{GANNT} 
        & T-SimCSE     & 0.25 & 0.49 & 0.33 & 0.41 \\
        & LiSSA-CoT-GPT4o   & \textbf{0.61} & 0.54 & \textbf{0.57} & \textbf{0.56} \\
        & LiSSA-CoT-Llama3.1 & 0.57 & 0.53 & 0.55 & 0.54 \\
        & WQI          & 0.26 & 0.56 & 0.35 & 0.45 \\
        & S2Trace      & 0.29 & 0.54 & 0.38 & 0.46 \\
        & GeT2Trace    & 0.31 & \textbf{0.57} & 0.40 & 0.49 \\
    \midrule
    \multirow{6}{*}{CM1}
        & T-SimCSE     & 0.18 & \textbf{0.76} & 0.30 & 0.47 \\
        & LiSSA-CoT-GPT4o   & \textbf{0.46} & 0.60 & \textbf{0.52} & 0.57 \\
        & LiSSA-CoT-Llama3.1 & 0.41 & 0.64 & 0.50 & \textbf{0.58} \\
        & WQI          & 0.37 & 0.33 & 0.35 & 0.34 \\
        & S2Trace      & 0.31 & 0.48 & 0.38 & 0.44 \\
        & GeT2Trace    & 0.32 & 0.50 & 0.39 & 0.45 \\
    \midrule
    \multirow{3}{*}{CCHIT}
        & T-SimCSE     & 0.16 & \textbf{0.21} & 0.18 & \textbf{0.19 }\\
        & LiSSA-CoT-GPT4o   & \textbf{0.37} & 0.14 & \textbf{0.20} & 0.16 \\
        & LiSSA-CoT-Llama3.1 & 0.28 & 0.13 & 0.18 & 0.15 \\
    \midrule
    \multirow{5}{*}{MODIS}
        & T-SimCSE     & 0.24 & \textbf{0.68} & \textbf{0.35} & \textbf{0.50} \\
        & LiSSA-CoT-GPT4o   & \textbf{0.50} & 0.17 & 0.26 & 0.20 \\
        & LiSSA-CoT-Llama3.1 & 0.23 & 0.24 & 0.24 & 0.24 \\
    \midrule
    \multirow{3}{*}{Dronology}
        & T-SimCSE     & 0.18 &\textbf{ 0.70} & 0.29 & 0.44 \\
        & LiSSA-CoT-GPT4o   & 0.51 & 0.66 & \textbf{0.58} & \textbf{0.62} \\
        & LiSSA-CoT-Llama3.1 & 0.41 & 0.57 & 0.47 & 0.53 \\
    \midrule
    \multirow{3}{*}{WARC}
        & T-SimCSE     & 0.20 & \textbf{0.73} & 0.31 & 0.48 \\
        & LiSSA-CoT-GPT4o   & \textbf{0.54} & 0.64 & \textbf{0.58} & \textbf{0.62} \\
        & LiSSA-CoT-Llama3.1 & 0.43 & 0.64 & 0.52 & 0.58 \\
    \bottomrule
    \end{tabular}
\end{table}

As shown in Table~\ref{tab4}, across seven datasets with prior approaches (EasyClinic, GANNT, CM1, CCHIT, MODIS, Dronology, WARC), T-SimCSE attains the highest Recall on five (CM1, CCHIT, MODIS, Dronology, WARC). 
Correspondingly, T-SimCSE achieves the best $F_2$ on CCHIT and MODIS, and the best $F_1$ on EasyClinic and MODIS.

On GANNT, CM1, CCHIT, MODIS, Dronology, and WARC, we compared T-SimCSE, LiSSA-CoT-GPT4o, and LiSSA-CoT-Llama3.1 using the Friedman test on both Precision and Recall. The Precision differences were significant ($\chi^{2}(2)=10.33$, $p=0.0057$). Nemenyi’s post-hoc test showed that LiSSA-CoT-GPT4o achieved a significantly better average rank than T-SimCSE ($\lvert \text{rank diff} \rvert = 1.833 > \mathrm{CD} = 1.353$), while other pairwise contrasts were not significant. For Recall, the overall differences were not significant ($\chi^{2}(2)=4.08$, $p=0.130$), and no pairwise contrast reached significance.

From the perspective of $F_1$ and $F_2$ results, the CoT-based approaches achieve a better balance between precision and recall. Specifically, LiSSA-CoT-GPT4o outperforms T-SimCSE on $F_1$ in 5 out of 6 datasets and on $F_2$ in 4 out of 6 datasets; LiSSA-CoT-Llama3.1 likewise surpasses T-SimCSE on $F_1$ in 5 out of 6 datasets and on $F_2$ in 4 out of 6 datasets. This indicates that although T-SimCSE often has an advantage in recall, the CoT-based approaches, by offering a more balanced trade-off between precision and recall, demonstrate superior overall effectiveness.

Further analysis of WQI, S2Trace, and GeT2Trace shows that on EasyClinic, all three have lower Precision and $F_1$ than T-SimCSE. On GANNT, T-SimCSE underperforms across metrics. On CM1, the three approaches achieve higher $F_1$, whereas T-SimCSE attains the highest Recall and the best $F_2$, indicating a recall-oriented operating point.

Overall, while T-SimCSE does not match the performance of CoT-based approaches, it still shows advantages on certain datasets and remains competitive against non-LLM baselines, particularly in recall-oriented scenarios.

\subsection{RQ4: Is the rewarding strategy for the TAs effective?}\label{subsec6.4}

\begin{sidewaystable}[htbp]
\centering
\small
\caption{Results across different approaches and datasets}
\label{tab:results3}
\begingroup
\setlength{\tabcolsep}{1.2pt}
\renewcommand{\arraystretch}{1.12}

\begin{tabular}{@{}l*{12}{c}@{}}
\toprule
\multirow{2}{*}{Model}
& \multicolumn{4}{c}{EasyClinic}
& \multicolumn{4}{c}{GANNT}
& \multicolumn{4}{c}{CM1} \\
\cmidrule(lr){2-5}\cmidrule(lr){6-9}\cmidrule(lr){10-13}
& {$k_1$/$k_2$} & MAP & $p$ & $\delta$
& {$k_1$/$k_2$} & MAP & $p$ & $\delta$
& {$k_1$/$k_2$} & MAP & $p$ & $\delta$ \\
\midrule
T-SimCSE               & 0.07/0.04 & 0.81 & --   & --   & 0.03/0.09 & 0.51 & --   & --   & 0.02/0.18 & 0.47 & --   & --   \\
SimCSE                 & --        & 0.77 & 0.01 & 0.34 & --        & 0.47 & 0.00 & 0.24 & --        & 0.46 & 0.00 & 0.20 \\
Word2Vec               & 0.07/0.10 & 0.74 & --   & --   & 0.11/0.91 & 0.32 & --   & --   & 0.01/0.03 & 0.44 & --   & --   \\
Word2Vec-NR            & --        & 0.71 & 0.01 & 0.38 & --        & 0.29 & 0.03 & 0.18 & --        & 0.41 & 0.03 & 0.26 \\
BERT                   & 0.04/0.03 & 0.53 & --   & --   & 0.03/0.06 & 0.32 & --   & --   & 0.17/0.16 & 0.31 & --   & --   \\
BERT-NR                & --        & 0.50 & 0.11 & 0.38 & --        & 0.30 & 0.00 & 0.26 & --        & 0.29 & 0.02 & 0.10 \\
LSI                    & 0.06/0.05 & 0.75 & --   & --   & 0.02/0.06 & 0.46 & --   & --   & 0.16/0.15 & 0.70 & --   & --   \\
LSI-NR                 & --        & 0.73 & 0.01 & 0.80 & --        & 0.41 & 0.00 & 0.26 & --        & 0.67 & 0.00 & 0.42 \\
VSM                    & 0.20/0.46 & 0.72 & --   & --   & 0.02/0.04 & 0.39 & --   & --   & 0.14/0.11 & 0.62 & --   & --   \\
VSM-NR                 & --        & 0.70 & 0.03 & 0.34 & --        & 0.34 & 0.01 & 0.26 & --        & 0.57 & 0.00 & 0.50 \\
\end{tabular}

\vspace{4pt}

\begin{tabular}{@{}l*{12}{c}@{}}
\multirow{2}{*}{Model}
& \multicolumn{4}{c}{CCHIT}
& \multicolumn{4}{c}{MODIS}
& \multicolumn{4}{c}{Dronology} \\
\cmidrule(lr){2-5}\cmidrule(lr){6-9}\cmidrule(lr){10-13}
& {$k_1$/$k_2$} & MAP & $p$ & $\delta$
& {$k_1$/$k_2$} & MAP & $p$ & $\delta$
& {$k_1$/$k_2$} & MAP & $p$ & $\delta$ \\
\midrule
T-SimCSE               & 0.01/0.01 & 0.22 & --   & --   & 0.43/0.46 & 0.74 & --   & --   & 0.01/0.03 & 0.60 & --   & --   \\
SimCSE                 & --        & 0.22 & 0.13 & 0.06 & --        & 0.66 & 0.00 & 0.76 & --        & 0.59 & 0.05 & 0.12 \\
Word2Vec               & 0.82/0.93 & 0.17 & --   & --   & 0.49/0.30 & 0.62 & --   & --   & 0.11/0.08 & 0.47 & --   & --   \\
Word2Vec-NR            & --        & 0.16 & 0.00 & 0.12 & --        & 0.44 & 0.00 & 0.88 & --        & 0.47 & 1.00 & 0.00 \\
BERT                   & 0.35/0.01 & 0.08 & --   & --   & 0.05/0.07 & 0.54 & --   & --   & 0.16/0.90 & 0.41 & --   & --   \\
BERT-NR                & --        & 0.08 & 0.00 & 0.08 & --        & 0.46 & 0.00 & 1.00 & --        & 0.40 & 0.00 & 0.14 \\
LSI                    & 0.09/0.19 & 0.19 & --   & --   & 0.67/0.79 & 0.56 & --   & --   & 0.98/0.02 & 0.57 & --   & --   \\
LSI-NR                 & --        & 0.19 & 0.06 & 0.06 & --        & 0.39 & 0.00 & 0.58 & --        & 0.56 & 0.00 & 0.20 \\
VSM                    & 0.02/0.94 & 0.19 & --   & --   & 0.01/0.02 & 0.57 & --   & --   & 0.01/0.02 & 0.54 & --   & --   \\
VSM-NR                 & --        & 0.18 & 0.01 & 0.06 & --        & 0.27 & 0.00 & 1.00 & --        & 0.52 & 0.01 & 0.20 \\
\end{tabular}

\vspace{4pt}

\begin{tabular}{@{}l*{16}{c}@{}}
\multirow{2}{*}{Model}
& \multicolumn{4}{c}{WARC}
& \multicolumn{4}{c}{InfusionPump}
& \multicolumn{4}{c}{EBT}
& \multicolumn{4}{c}{IceBreaker} \\
\cmidrule(lr){2-5}\cmidrule(lr){6-9}\cmidrule(lr){10-13}\cmidrule(lr){14-17}
& {$k_1$/$k_2$} & MAP & $p$ & $\delta$
& {$k_1$/$k_2$} & MAP & $p$ & $\delta$
& {$k_1$/$k_2$} & MAP & $p$ & $\delta$
& {$k_1$/$k_2$} & MAP & $p$ & $\delta$ \\
\midrule
T-SimCSE               & 0.01/0.02 & 0.65 & --   & --   & 0.01/0.04 & 0.42 & --   & --   & 0.05/0.22 & 0.83 & --   & --   & 0.03/0.06 & 0.45 & --   & --   \\
SimCSE                 & --        & 0.62 & 0.00 & 0.30 & --        & 0.42 & 0.70 & -0.02 & --        & 0.80 & 0.00 & 0.50 & --        & 0.44 & 0.04 & 0.06 \\
Word2Vec               & 0.06/0.10 & 0.48 & --   & --   & 0.15/0.03 & 0.43 & --   & --   & 0.01/0.13 & 0.60 & --   & --   & 0.11/0.10 & 0.41 & --   & --   \\
Word2Vec-NR            & --        & 0.39 & 0.00 & 0.26 & --        & 0.39 & 0.00 & 0.26 & --        & 0.16 & 0.00 & 1.00 & --        & 0.40 & 0.01 & 0.08 \\
BERT                   & 0.07/0.04 & 0.49 & --   & --   & 0.15/0.12 & 0.27 & --   & --   & 0.01/0.86 & 0.64 & --   & --   & 0.03/0.04 & 0.26 & --   & --   \\
BERT-NR                & --        & 0.48 & 0.01 & 0.20 & --        & 0.26 & 1.00 & -0.12& --        & 0.62 & 0.23 & 0.00 & --        & 0.26 & 0.16 & 0.06 \\
LSI                    & 0.06/0.08 & 0.62 & --   & --   & 0.03/0.01 & 0.41 & --   & --   & 0.05/0.17 & 0.68 & --   & --   & 0.02/0.07 & 0.40 & --   & --   \\
LSI-NR                 & --        & 0.59 & 0.00 & 0.26 & --        & 0.38 & 0.00 & 0.30 & --        & 0.66 & 0.00 & 1.00 & --        & 0.39 & 0.04 & 0.06 \\
VSM                    & 0.01/0.02 & 0.56 & --   & --   & 0.01/0.04 & 0.41 & --   & --   & 0.09/0.48 & 0.62 & --   & --   & 0.01/0.03 & 0.39 & --   & --   \\
VSM-NR                 & --        & 0.55 & 0.00 & 0.16 & --        & 0.38 & 0.00 & 0.32 & --        & 0.61 & 0.00 & 1.00 & --        & 0.38 & 0.11 & 0.06 \\
\bottomrule
\end{tabular}

\par\medskip
\noindent\makebox[\linewidth][l]{\footnotesize\textit{NR = none-reordering variant.}}
\endgroup
\end{sidewaystable}

\begin{figure}[t]
\centering

\begin{subfigure}[b]{0.20\textwidth}
    \centering
    \includegraphics[width=\linewidth]{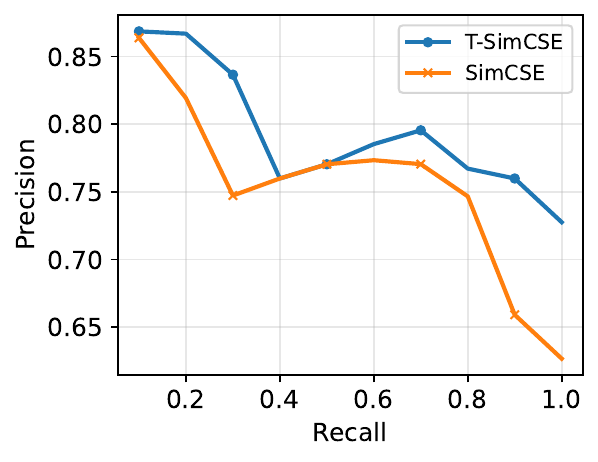}
    \caption{EasyClinic}
    \label{fig5:1}
\end{subfigure}\hfill
\begin{subfigure}[b]{0.20\textwidth}
    \centering
    \includegraphics[width=\linewidth]{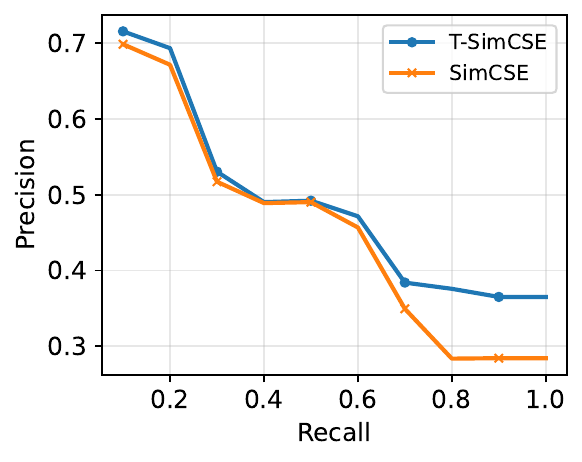}
    \caption{GANNT}
    \label{fig5:2}
\end{subfigure}\hfill
\begin{subfigure}[b]{0.20\textwidth}
    \centering
    \includegraphics[width=\linewidth]{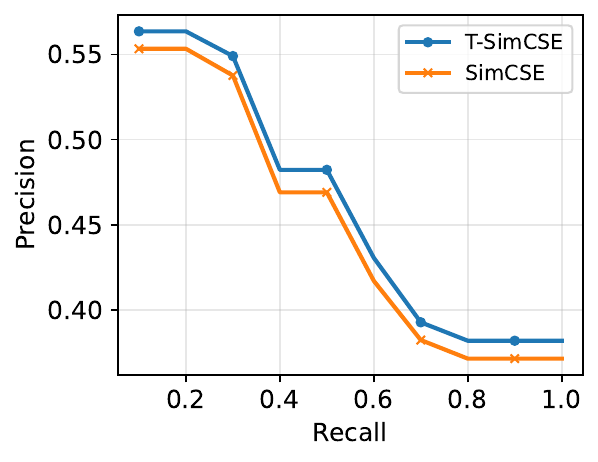}
    \caption{CM1}
    \label{fig5:3}
\end{subfigure}\hfill
\begin{subfigure}[b]{0.20\textwidth}
    \centering
    \includegraphics[width=\linewidth]{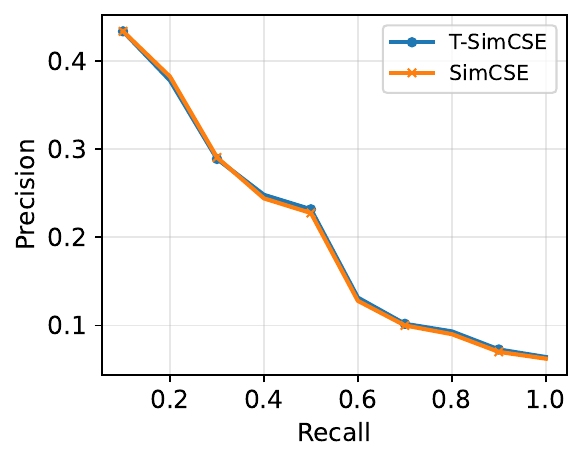}
    \caption{CCHIT}
    \label{fig5:4}
\end{subfigure}\hfill
\begin{subfigure}[b]{0.20\textwidth}
    \centering
    \includegraphics[width=\linewidth]{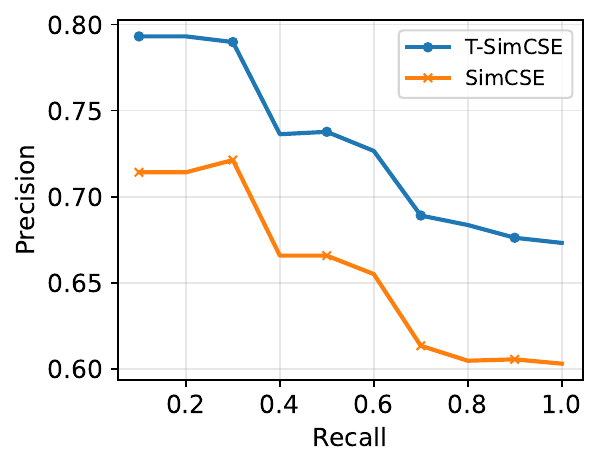}
    \caption{MODIS}
    \label{fig5:5}
\end{subfigure}

\vspace{0.1em}

\begin{subfigure}[b]{0.20\textwidth}
    \centering
    \includegraphics[width=\linewidth]{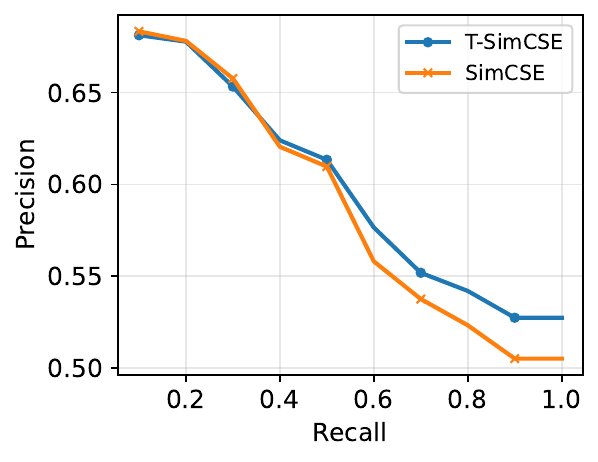}
    \caption{Dronology}
    \label{fig5:6}
\end{subfigure}\hfill
\begin{subfigure}[b]{0.20\textwidth}
    \centering
    \includegraphics[width=\linewidth]{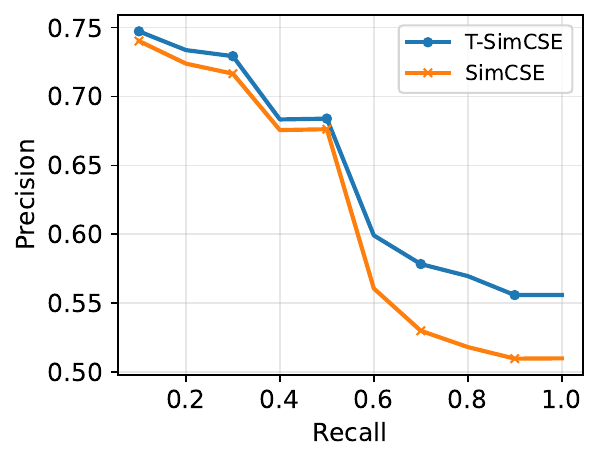}
    \caption{WARC}
    \label{fig5:7}
\end{subfigure}\hfill
\begin{subfigure}[b]{0.20\textwidth}
    \centering
    \includegraphics[width=\linewidth]{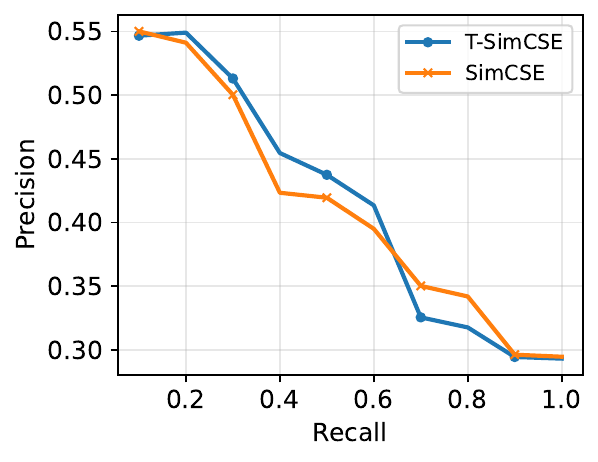}
    \caption{InfusionPump}
    \label{fig5:8}
\end{subfigure}\hfill
\begin{subfigure}[b]{0.20\textwidth}
    \centering
    \includegraphics[width=\linewidth]{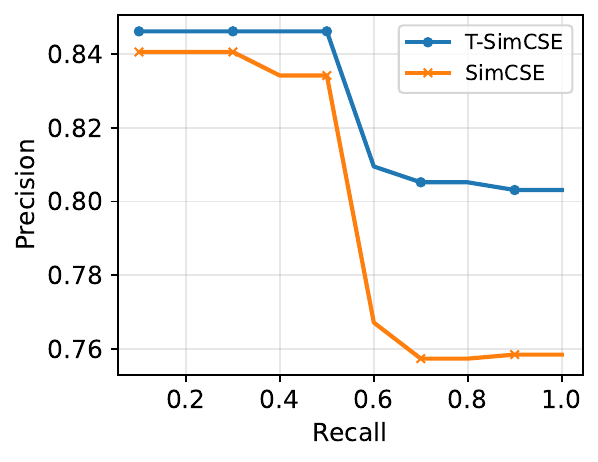}
    \caption{EBT}
    \label{fig5:9}
\end{subfigure}\hfill
\begin{subfigure}[b]{0.20\textwidth}
    \centering
    \includegraphics[width=\linewidth]{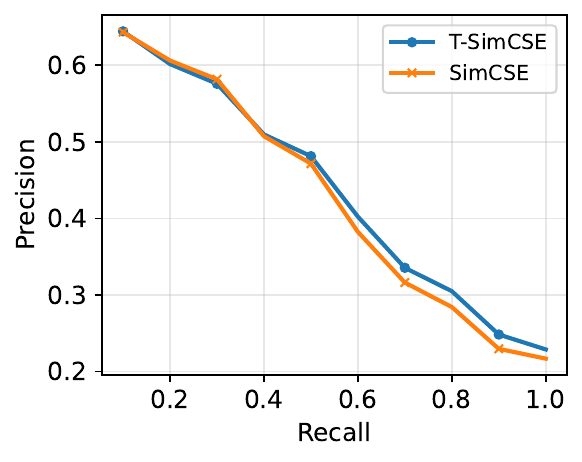}
    \caption{IceBreaker}
    \label{fig5:10}
\end{subfigure}

\caption{The comparison of the Precision-Recall curves of T-SimCSE and SimCSE}
\label{fig5}
\end{figure}

To address RQ4, we conducted an ablation study to isolate the impact of the rewarding strategy: across the ten datasets, we compared the approach with the strategy enabled versus removed and performed Wilcoxon signed-rank tests using the precision values at recall levels from 0.1 to 1.0. As shown in Table~\ref{tab:results3} and Figure~\ref{fig5}, enabling the rewarding strategy increases MAP, strengthens the statistical evidence, and improves the Precision–Recall curves; the gains vary by dataset, ranging from small to large.

As shown in Table~\ref{tab:results3}, across ten datasets and multiple baselines, the reordering module yields statistically significant improvements on MODIS and EBT with large effect sizes, and on EasyClinic for most baselines (BERT being borderline), while on CM1 all comparisons are significant with effect sizes ranging from small to large (notably medium/large for LSI and VSM). On GANNT and WARC, it delivers statistically significant but small gains; on Dronology, IceBreaker, and CCHIT, the effect is negligible or borderline. For InfusionPump with the BERT and T-SimCSE baselines,  Wilcoxon signed-rank tests indicate a slight disadvantage or no difference. Consistent with these findings, in terms of MAP, the reordering variant outperforms the none-reordering counterpart in $48/50$ paired comparisons (ties in $2/50$). Overall, reordering provides a stable enhancement to traceability link recovery quality, with both statistical significance and practically meaningful effect sizes on several datasets.

\begin{figure}[ht]
\centering
\includegraphics[width=0.8\textwidth]{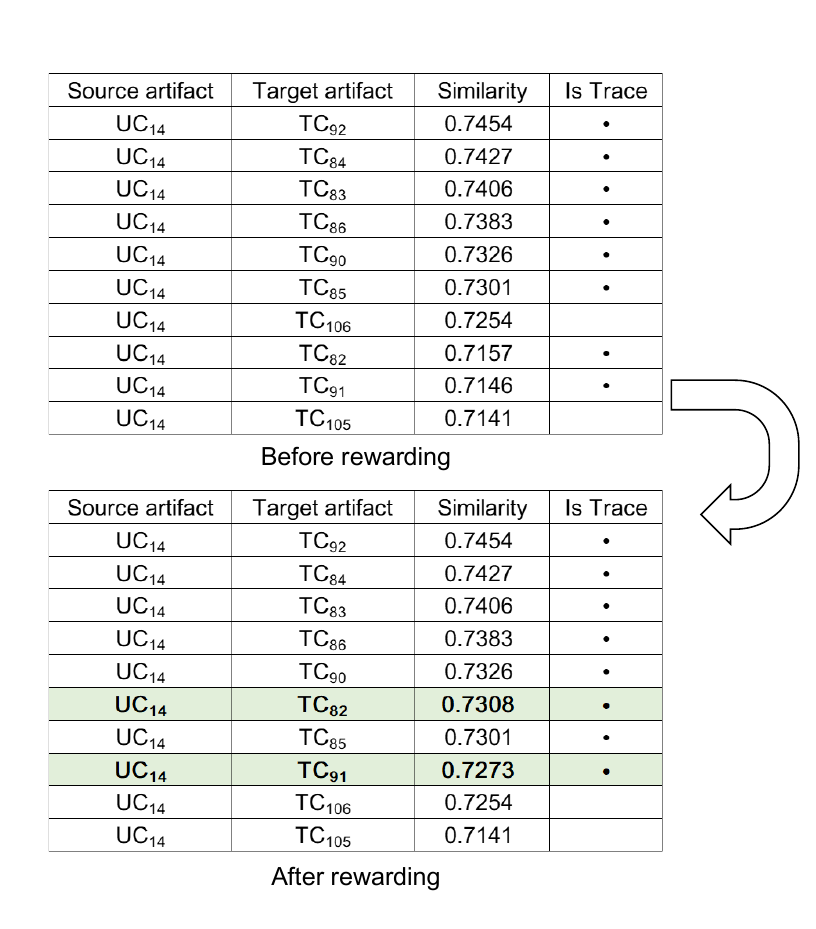}
\caption{The ranking of TAs before and after rewarding}
\label{fig6}
\end{figure}

These experimental findings demonstrate that the rewarding strategy does find TAs with trace links to the SAs via HPTAs, thus further improving the performance of recovering trace links. To demonstrate the effect of the rewarding strategy more clearly, we use $UC_{14}$ in the EasyClinic dataset as an example. Figure~\ref{fig6} shows the top 10 TAs before and after rewarding, respectively. The results show that the 8th-ranked $TC_{82}$ and the 9th-ranked $TC_{91}$ trace links are ranked 6th and 8th respectively after being rewarded.

\section{Implications}\label{sec7}

To recover trace links between requirements and other artifacts, the most common way is to compare the semantic similarity between the SAs and TAs, and then create trace links between the higher-scoring artifact pairs. However, such approaches face a challenge in that they ignore the impact of grammar on the accuracy of semantic expressions in natural language, especially the absence of word order information~\citet{chen2021}. As a result, researchers began to rely on DL approaches to solve the problem. Training DL models requires a large amount of project data. Yet, collecting and saving project data in software development is not given too much attention, which becomes another challenge.

The experimental results demonstrate that T-SimCSE addresses these challenges through two key design considerations. The first consideration is the selection of the PLM---SimCSE. SimCSE integrates BERT with contrastive learning objectives, allowing it to perform extremely well in the sentence embedding task. The second consideration is the rewarding strategy integrating specificity of TAs, which further improves the accuracy of the trace links. Although there has been some research on applying rewarding strategy to the trace links recovery, most of them focus on structured documents such as source code. The rewarding strategy in T-SimCSE focuses on NL TAs (e.g. requirements, test cases, and so on). The definition of specificity argues that not all semantically close artifacts are equally important and that they should be differentiated according to their specificity.
Although LLM-based approaches generally achieve stronger overall performance, T-SimCSE still demonstrates competitive results in scenarios where large models cannot be used and on several datasets where it performs comparatively well. However, in scenarios where LLMs can be utilized, future work could explore more LLM-based approaches to further enhance performance.

The proposed rewarding strategy can be applied not only to the traceability link recovery, but also to various recommendation tasks such as API recommendations for constructing composite services. In future work, we consider applying this rewarding strategy to other recommendation tasks and observing the results.

\textbf{Practical applicability:} While T-SimCSE is technically a fully automated approach in terms of execution (i.e., once artifacts are uploaded, links are automatically generated), the current precision and recall values are not yet sufficient to enable reliable full automation in industrial practice. Instead, the approach is more appropriate for a \textit{semi-automated scenario}, where requirement analysts use the ranked candidate links provided by T-SimCSE as guidance. In such settings, analysts only need to review a small top-$k$ subset of candidate links rather than exhaustively inspecting all possible artifact pairs, which can substantially reduce the manual effort. 

\section{Threats to Validity}\label{sec8}

In this section, we provide a detailed discussion of the validity threats associated with our approach.

\subsection{Threats to internal validity}\label{subsec8.1}

Threats to internal validity are primarily related to experimental errors. To ensure an unbiased experimental process, the code was carefully reviewed by two developers. We modified the official open-source code of the baseline approaches and adapted it to match the required input and output formats for our experiments. Therefore, the likelihood of threats arising from the implementation of the approaches is minimal.

\subsection{Threats to external validity}\label{subsec8.2}

Threats to external validity primarily stem from parameter optimization and dataset selection.

First, our approach demands high accuracy from the learning model. This requirement stems from the integration of similarity computation with task specificity, as the rewarding strategy's correctness relies on precise model performance. Notably, in this study, we used the default parameters of the pre-trained SimCSE model, as its current accuracy was deemed satisfactory. In future work, we plan to further enhance the model by fine-tuning its parameters.

Second, we selected ten commonly used datasets in the requirements traceability task, covering artifacts such as use cases, test cases, high-level requirements, low-level requirements, design definitions, and class diagrams descriptions. However, these datasets exhibit limitations in representing real industrial scenarios.  

\begin{itemize}
    \item Scale and structure mismatch: The datasets are relatively small and fragmented, whereas industrial environments often involve long documents that pose challenges in segmentation granularity.
    \item Incomplete artifact coverage: Real-world projects include additional artifacts (e.g., Architecture Decision Records, detailed test protocols) whose compatibility with T-SimCSE remains unverified.
    \item Domain specificity: The current datasets are confined to narrow domains and lack comprehensive coverage of the software engineering landscape, particularly in multi-domain industrial contexts.
\end{itemize}

To address these gaps, future work will extend T-SimCSE's applicability by integrating datasets that:  
\begin{itemize}
    \item Encompass a broader variety of artifacts (e.g., regulatory documents, multilingual specifications)
    \item Better reflect industrial-scale document structures and multi-domain characteristics.
\end{itemize}

\subsection{Threats to construct validity}\label{subsec8.3}

The threat to construct validity is related to the accuracy of the evaluation metrics used in the experiment. Precision, Recall, $F_1$, $F_2$ and MAP are commonly used evaluation metrics in the field of traceability link recovery, and have been widely applied in previous studies.

\section{Conclusions}\label{sec9}

This paper presents T-SimCSE, a fully automated requirements traceability link recovery
approach based on the PLM—--SimCSE---by integrating with a rewarding strategy. T-SimCSE is designed to improve the accuracy of requirements trace links and overcome
the problem that the large amount of training data is required by most DL-based
approaches. It automates the generation of trace links by enabling a SimCSE model
and a rewarding strategy based on the similarity and specificity. To validate the effectiveness of T-SimCSE, we formulated four research questions and evaluated them on ten public datasets. The results show that our approach outperforms the baseline approaches BERT-based, Word2Vec-based, VSM-based and LSI-based in terms of precision, recall, and MAP. In addition, we compared our approach with several prior approaches, including S2Trace, WQI, LiSSA-CoT-GPT4o, and GeT2Trace. On certain datasets, our approach achieves higher scores than these prior approaches in certain metrics among  $F_1$, $F_2$, precision, and recall. We also discuss whether the rewarding strategy is helpful through the ablation experiments. In the future, T-SimCSE can be improved from three aspects, which includes: 1) mining more implicit information between TAs as criteria for rewards; 2) applying T-SimCSE to real projects to evaluate its feasibility; 3) combining T-SimCSE with LLMs such as GPT-4 to further improve its performance. Although T-SimCSE is now fully automated by batch uploading artifacts in the form of both xml and txt, it is necessary to support more formats of artifacts and develop an user interface for requirement analysts to facilitate with its application.

\bibliography{sn-bibliography}

@article{DellAydeDalp2023,
  author  = {Dell'Anna, Davide and Aydemir, Fatma Ba\c{s}ak and Dalpiaz, Fabiano},
  title   = {Evaluating Classifiers in {SE} Research: The {ECSER} Pipeline and Two Replication Studies},
  journal = {Empirical Software Engineering},
  year    = {2023},
  volume  = {28},
  number  = {3},
  doi     = {10.1007/s10664-022-10243-1}
}

@inproceedings{bib2,
  author    = {Gotel, O. and Cleland-Huang, J. and Hayes, Jane Huffman and Zisman, A. and Egyed, A. and Gr{\"u}nbacher, P. and Antoniol, G.},
  title     = {The Quest for Ubiquity: A Roadmap for Software and Systems Traceability Research},
  booktitle = {20th IEEE International Requirements Engineering Conference (RE)},
  year      = {2012},
  pages     = {71--80},
  doi       = {10.1109/RE.2012.6345841}
}

@incollection{gotel2012traceability,
  author    = {Gotel, Orlena and Cleland-Huang, Jane and Hayes, Jane Huffman and Zisman, Andrea and Egyed, Alexander and Gr{\"u}nbacher, Paul and Dekhtyar, Alex and Antoniol, Giuliano and Maletic, Jonathan and M{\"a}der, Patrick},
  title     = {Traceability Fundamentals},
  booktitle = {Software and Systems Traceability},
  year      = {2012},
  publisher = {Springer London},
  address   = {London},
  pages     = {3--22},
  isbn      = {978-1-4471-2239-5},
  doi       = {10.1007/978-1-4471-2239-5_1}
}

@book{lamsweerde2009requirements,
  title={Requirements engineering: from system goals to UML models to software specifications},
  author={van Lamsweerde, Axel},
  year={2009},
  publisher={John Wiley \& Sons, Ltd}
}

@article{sundaram2010assessing,
  title={Assessing traceability of software engineering artifacts},
  author={Sundaram, Senthil Karthikeyan and Hayes, Jane Huffman and Dekhtyar, Alex and Holbrook, E Ashlee},
  journal={Requirements engineering},
  volume={15},
  pages={313--335},
  year={2010},
  publisher={Springer},
  doi={10.1007/s00766-009-0096-6}
}

@inproceedings{bib6,
  author={Kuang, Hongyu and Mäder, Patrick and Hu, Hao and Ghabi, Achraf and Huang, LiGuo and Jian, Lv and Egyed, Alexander},
  booktitle={2012 28th IEEE International Conference on Software Maintenance (ICSM)}, 
  title={Do data dependencies in source code complement call dependencies for understanding requirements traceability?}, 
  year={2012},
  pages={181-190},
  doi={10.1109/ICSM.2012.6405270}
}

@inproceedings{bib7,
  author={Kuang, Hongyu and Nie, Jia and Hu, Hao and Rempel, Patrick and Lü, Jian and Egyed, Alexander and Mäder, Patrick},
  booktitle={2017 IEEE 24th International Conference on Software Analysis, Evolution and Reengineering (SANER)}, 
  title={Analyzing closeness of code dependencies for improving IR-based Traceability Recovery}, 
  year={2017},
  pages={68-78},
  doi={10.1109/SANER.2017.7884610}
}

@inproceedings{bib8,
  author={Goodrum, Micayla and Cleland-Huang, Jane and Lutz, Robyn and Cheng, Jinghui and Metoyer, Ronald},
  booktitle={2017 IEEE 25th International Requirements Engineering Conference (RE)}, 
  title={What Requirements Knowledge Do Developers Need to Manage Change in Safety-Critical Systems?}, 
  year={2017},
  pages={90-99},
  doi={10.1109/RE.2017.65}
}

@inproceedings{guo2017semantically,
  author={Guo, Jin and Cheng, Jinghui and Cleland-Huang, Jane},
  booktitle={2017 IEEE/ACM 39th International Conference on Software Engineering (ICSE)}, 
  title={Semantically Enhanced Software Traceability Using Deep Learning Techniques}, 
  year={2017},
  volume={},
  number={},
  pages={3-14},
  doi={10.1109/ICSE.2017.9}
}

@misc{lin2022enhancing,
  title={Enhancing Automated Software Traceability by Transfer Learning from Open-World Data}, 
  author={Jinfeng Lin and Amrit Poudel and Wenhao Yu and Qingkai Zeng and Meng Jiang and Jane Cleland-Huang},
  year={2022},
  eprint={2207.01084},
  archivePrefix={arXiv},
  primaryClass={cs.SE},
  url={https://arxiv.org/abs/2207.01084}, 
}

@inproceedings{rodriguez2023prompts,
  author={Rodriguez, Alberto D. and Dearstyne, Katherine R. and Cleland-Huang, Jane},
  booktitle={2023 IEEE 31st International Requirements Engineering Conference Workshops (REW)}, 
  title={Prompts Matter: Insights and Strategies for Prompt Engineering in Automated Software Traceability}, 
  year={2023},
  pages={455-464},
  doi={10.1109/REW57809.2023.00087}
}

@article{antoniol2002recovering,
  title={Recovering traceability links between code and documentation},
  author={Antoniol, Giuliano and Canfora, Gerardo and Casazza, Gregorio and Lucia, Andrea De and Merlo, Ettore},
  journal={IEEE Transactions on Software Engineering},
  volume={28},
  number={10},
  pages={970--983},
  month={Oct},
  year={2002},
  publisher={IEEE},
  doi={10.1109/TSE.2002.1041053}
}

@inproceedings{bib10,
  author={Marcus, A. and Maletic, J.I.},
  booktitle={25th International Conference on Software Engineering}, 
  title={Recovering documentation-to-source-code traceability links using latent semantic indexing}, 
  year={2003},
  volume={},
  number={},
  pages={125-135},
  doi={10.1109/ICSE.2003.1201194}
}

@inproceedings{bib11,
  author={Abadi, Aharon and Nisenson, Mordechai and Simionovici, Yahalomit},
  booktitle={2008 16th IEEE International Conference on Program Comprehension}, 
  title={A Traceability Technique for Specifications}, 
  year={2008},
  volume={},
  number={},
  pages={103-112},
  doi={10.1109/ICPC.2008.30}
}

@article{bib12,
  author={Guo, Jin and Gibiec, Marek and Cleland-Huang, Jane},
  title={Tackling the term-mismatch problem in automated trace retrieval},
  journal={Empirical Software Engineering},
  volume={22},
  pages={1103--1142},
  year={2017},
  doi={10.1007/s10664-016-9479-8}
}

@inproceedings{gao2021simcse,
  title={SimCSE: Simple Contrastive Learning of Sentence Embeddings},
  author={Gao, Tianyu and Yao, Xingcheng and Chen, Danqi},
  booktitle={2021 Conference on Empirical Methods in Natural Language Processing, EMNLP 2021},
  pages={6894--6910},
  year={2021},
  url={http://arks.princeton.edu/ark:/88435/pr1z892f7x}
}

@article{chen2021,
  title={Enhancing Requirements Traceability Recovery via a Graph Mining-Based Expansion Learning},
  author={Chen Lei and Wang Dandan and Wang Qing and Shi Lin},
  journal={Journal of Computer Research and Development},
  volume={58},
  number={4},
  pages={777--793},
  year={2021},
  issn = {1000-1239},
  doi={10.7544/issn1000-1239.2021.20200733}
}

@inproceedings{gethers2011integrating,
  author={Gethers, Malcom and Oliveto, Rocco and Poshyvanyk, Denys and Lucia, Andrea De},
  booktitle={2011 27th IEEE International Conference on Software Maintenance (ICSM)}, 
  title={On integrating orthogonal information retrieval methods to improve traceability recovery}, 
  year={2011},
  volume={},
  number={},
  pages={133-142},
  doi={10.1109/ICSM.2011.6080780}
}

@inproceedings{mills2019tracing,
  author={Mills, Chris and Escobar-Avila, Javier and Bhattacharya, Aditya and Kondyukov, Grigoriy and Chakraborty, Shayok and Haiduc, Sonia},
  booktitle={2019 IEEE International Conference on Software Maintenance and Evolution (ICSME)}, 
  title={Tracing with Less Data: Active Learning for Classification-Based Traceability Link Recovery}, 
  year={2019},
  volume={},
  number={},
  pages={103-113},
  doi={10.1109/ICSME.2019.00020}
}

@article{mahmoud2016detecting,
  title={Detecting, classifying, and tracing non-functional software requirements},
  author={Mahmoud, Anas and Williams, Grant},
  journal={Requirements Engineering},
  volume={21},
  pages={357-381},
  year={2016},
  doi={10.1007/s00766-016-0252-8}
}

@inproceedings{rath2018traceability,
author = {Rath, Michael and Rendall, Jacob and Guo, Jin L. C. and Cleland-Huang, Jane and M\"{a}der, Patrick},
title = {Traceability in the wild: automatically augmenting incomplete trace links},
year = {2018},
isbn = {9781450356381},
publisher = {Association for Computing Machinery},
address = {New York, NY, USA},
doi = {10.1145/3180155.3180207},
booktitle = {Proceedings of the 40th International Conference on Software Engineering},
pages = {834-845},
numpages = {12},
location = {Gothenburg, Sweden},
}

@inproceedings{moran2020improving,
author = {Moran, Kevin and Palacio, David N. and Bernal-C\'{a}rdenas, Carlos and McCrystal, Daniel and Poshyvanyk, Denys and Shenefiel, Chris and Johnson, Jeff},
title = {Improving the effectiveness of traceability link recovery using hierarchical bayesian networks},
year = {2020},
isbn = {9781450371216},
publisher = {Association for Computing Machinery},
address = {New York, NY, USA},
doi = {10.1145/3377811.3380418},
booktitle = {Proceedings of the ACM/IEEE 42nd International Conference on Software Engineering},
pages = {873-885},
numpages = {13},
location = {Seoul, South Korea},
}

@article{marcen2020traceability,
  title = {Traceability Link Recovery between Requirements and Models using an Evolutionary Algorithm Guided by a Learning to Rank Algorithm: Train control and management case},
  journal = {Journal of Systems and Software},
  volume = {163},
  pages = {110519},
  year = {2020},
  issn = {0164-1212},
  doi = {10.1016/j.jss.2020.110519},
  author = {Ana C. Marcén and Raúl Lapeña and Óscar Pastor and Carlos Cetina},
}

@inproceedings{liu2020towards,
  author={Liu, Yalin and Lin, Jinfeng and Zeng, Qingkai and Jiang, Meng and Cleland-Huang, Jane},
  booktitle={2020 IEEE 28th International Requirements Engineering Conference (RE)}, 
  title={Towards Semantically Guided Traceability}, 
  year={2020},
  volume={},
  number={},
  pages={328-333},
  doi={10.1109/RE48521.2020.00043}
}

@inproceedings{lin2021traceability,
 author={Lin, Jinfeng and Liu, Yalin and Zeng, Qingkai and Jiang, Meng and Cleland-Huang, Jane},
  booktitle={2021 IEEE/ACM 43rd International Conference on Software Engineering (ICSE)}, 
  title={Traceability Transformed: Generating More Accurate Links with Pre-Trained BERT Models}, 
  year={2021},
  volume={},
  number={},
  pages={324-335},
  doi={10.1109/ICSE43902.2021.00040}
}

@inproceedings{lam2015combining,
  author={Lam, An Ngoc and Nguyen, Anh Tuan and Nguyen, Hoan Anh and Nguyen, Tien N.},
  booktitle={2015 30th IEEE/ACM International Conference on Automated Software Engineering (ASE)}, 
  title={Combining Deep Learning with Information Retrieval to Localize Buggy Files for Bug Reports (N)}, 
  year={2015},
  volume={},
  number={},
  pages={476-481},
  doi={10.1109/ASE.2015.73},
  address={Lincoln},
}

@INPROCEEDINGS{mill2018automatic,
  author={Mills, Chris and Escobar-Avila, Javier and Haiduc, Sonia},
  booktitle={2018 IEEE International Conference on Software Maintenance and Evolution (ICSME)}, 
  title={Automatic Traceability Maintenance via Machine Learning Classification}, 
  year={2018},
  volume={},
  number={},
  pages={369-380},
  doi={10.1109/ICSME.2018.00045}}

@inproceedings{hey2021improving,
 author={Hey, Tobias and Chen, Fei and Weigelt, Sebastian and Tichy, Walter F.},
  booktitle={2021 IEEE International Conference on Software Maintenance and Evolution (ICSME)}, 
  title={Improving Traceability Link Recovery Using Fine-grained Requirements-to-Code Relations}, 
  year={2021},
  volume={},
  number={},
  pages={12-22},
  doi={10.1109/ICSME52107.2021.00008}
}

@inproceedings{gao2019representation,
  author       = {Jun Gao and
                  Di He and
                  Xu Tan and
                  Tao Qin and
                  Liwei Wang and
                  Tie{-}Yan Liu},
  title        = {Representation Degeneration Problem in Training Natural Language Generation
                  Models},
  booktitle    = {7th International Conference on Learning Representations},
  publisher    = {OpenReview.net},
  year         = {2019},
  url          = {https://openreview.net/forum?id=SkEYojRqtm},
}

@inproceedings{wang2019improving,
author       = {Lingxiao Wang and
                  Jing Huang and
                  Kevin Huang and
                  Ziniu Hu and
                  Guangtao Wang and
                  Quanquan Gu},
  title        = {Improving Neural Language Generation with Spectrum Control},
  booktitle    = {8th International Conference on Learning Representations},
  publisher    = {OpenReview.net},
  year         = {2020},
  url          = {https://openreview.net/forum?id=ByxY8CNtvr},
}

@inproceedings{wang2018enhancing,
  author={Wang, Wentao and Niu, Nan and Liu, Hui and Niu, Zhendong},
  booktitle={2018 IEEE 26th International Requirements Engineering Conference (RE)}, 
  title={Enhancing Automated Requirements Traceability by Resolving Polysemy}, 
  year={2018},
  volume={},
  number={},
  pages={40-51},
  doi={10.1109/RE.2018.00-53}
}

@inproceedings{Hayes2003Improving,
  author    = {Hayes, J. H. and Dekhtyar, A. and Osborne, J.},
  title     = {Improving requirements tracing via information retrieval},
  booktitle = {Proceedings of the 11th IEEE International Requirements Engineering Conference (ICRE 2003)},
  year      = {2003},
  pages     = {138--147},
  doi       = {10.1109/ICRE.2003.1232745}
}

@article{hayes2006Advancing,
  author    = {Hayes, J. H. and Dekhtyar, A. and Sundaram, S. K.},
  title     = {Advancing Candidate Link Generation for Requirements Tracing: The Study of Methods},
  journal   = {IEEE Transactions on Software Engineering},
  year      = {2006},
  volume    = {32},
  number    = {1},
  pages     = {4--19},
  doi       = {10.1109/TSE.2006.3}
}

@inproceedings{Lohar2013Improving,
  author    = {Lohar, S. and Amornborvornwong, S. and Zisman, A. and Cleland-Huang, J.},
  title     = {Improving Trace Accuracy Through Data-driven Configuration and Composition of Tracing Features},
  booktitle = {Proceedings of the 2013 9th Joint Meeting on Foundations of Software Engineering (ESEC/FSE 2013)},
  year      = {2013},
  pages     = {378--388},
  doi       = {10.1145/2491411.2491432}
}

@inproceedings{zhao2017improved,
  author    = {Zhao, T. and Cao, Q. and Sun, Q.},
  title     = {An Improved Approach to Traceability Recovery Based on Word Embeddings},
  booktitle = {Proceedings of the 24th Asia-Pacific Software Engineering Conference (APSEC 2017)},
  year      = {2017},
  pages     = {81--89},
  doi       = {10.1109/APSEC.2017.14}
}

@inproceedings{chen2019enhancing,
  author    = {Chen, L. and Wang, D. and Wang, J. and Wang, Q.},
  title     = {Enhancing Unsupervised Requirements Traceability with Sequential Semantics},
  booktitle = {Proceedings of the 26th Asia-Pacific Software Engineering Conference (APSEC 2019)},
  year      = {2019},
  pages     = {23--30},
  doi       = {10.1109/APSEC48747.2019.00013}
}

@incollection{Schlutter2021Improving,
  author    = {Schlutter, A. and Vogelsang, A.},
  title     = {Improving Trace Link Recovery Using Semantic Relation Graphs and Spreading Activation},
  booktitle = {Requirements Engineering: Foundation for Software Quality},
  year      = {2021},
  editor    = {Dalpiaz, F. and Spoletini, P.},
  pages     = {37--53},
  publisher = {Springer},
  doi       = {10.1007/978-3-030-73128-1_3}
}

@inproceedings{marcus2003recovering,
  title={Recovering Documentation-to-source-code Traceability Links Using Latent Semantic Indexing},
  author={Marcus, Andrian and Maletic, Jonathan I},
  booktitle={Proceedings of the 25th International Conference on Software Engineering},
  series={ICSE '03},
  pages={125--135},
  year={2003},
  organization={IEEE Computer Society},
  url={http://dl.acm.org/citation.cfm?id=776816}

}

@inproceedings{mahmoud2015information,
  title={An information theoretic approach for extracting and tracing non-functional requirements},
  author={Mahmoud, Ahmed},
  booktitle={2015 IEEE 23rd International Requirements Engineering Conference (RE)},
  pages={36--45},
  month={Aug},
  year={2015},
  doi={10.1109/RE.2015.7320406}
}

@inproceedings{asuncion2010software,
  title={Software traceability with topic modeling},
  author={Asuncion, Hazeline U and Asuncion, Arthur U and Taylor, Richard N},
  booktitle={2010 ACM/IEEE 32nd International Conference on Software Engineering},
  volume={1},
  pages={95--104},
  month={May},
  year={2010},
  doi={10.1145/1806799.1806817}
}

@inproceedings{gao2023using,
  title={Using Consensual Biterms from Text Structures of Requirements and Code to Improve IR-Based Traceability Recovery},
  author={Gao, Hui and Kuang, Hua and Sun, Kai and Ma, Xing and Egyed, Alfred and Mäder, Patrick and Rong, Guoping and Shao, Dong and Zhang, He},
  booktitle={Proceedings of the 37th IEEE/ACM International Conference on Automated Software Engineering},
  series={ASE '22},
  month={Jan},
  year={2023},
  organization={Association for Computing Machinery},
  doi={10.1145/3551349.3556948}
}

@inproceedings{devlin2019bert,
  title={Bert: Pre-training of deep bidirectional transformers for language understanding},
  author={Devlin, Jacob and Chang, Ming-Wei and Lee, Kenton and Toutanova, Kristina},
  booktitle={Proceedings of the 2019 conference of the North American chapter of the association for computational linguistics: human language technologies, volume 1 (long and short papers)},
  pages={4171--4186},
  year={2019}
}

@article{liu2019roberta,
  title={Roberta: A robustly optimized bert pretraining approach},
  author={Liu, Yinhan and Ott, Myle and Goyal, Naman and Du, Jingfei and Joshi, Mandar and Chen, Danqi and Levy, Omer and Lewis, Mike and Zettlemoyer, Luke and Stoyanov, Veselin},
  journal={arXiv preprint arXiv:1907.11692},
  year={2019}
}

@inproceedings{reimers2019sentencebert,
  title     = {Sentence-{BERT}: Sentence Embeddings Using Siamese {BERT}-Networks},
  author    = {Reimers, Nils and Gurevych, Iryna},
  booktitle = {Proceedings of the 2019 Conference on Empirical Methods in Natural Language Processing},
  year      = {2019},
  pages     = {3982--3992},
  publisher = {Association for Computational Linguistics},
  url       = {http://arxiv.org/abs/1908.10084}
}

@article{ni2022gtrt5,
  title   = {Large Dual Encoders Are Generalizable Retrievers},
  author  = {Xin Ni and Chen Wu and Hainan Zhang and Patrick Lewis and Wen-tau Yih and Sebastian Riedel and Fabio Petroni},
  journal = {Transactions of the Association for Computational Linguistics},
  year    = {2022},
  volume  = {10},
  pages   = {837--854},
  note    = {GTR-T5 family}
}

@article{wang2023e5,
  title   = {Text Embeddings by Weakly-Supervised Contrastive Learning},
  author  = {Yuqing Wang and Rio Maisyarah and Sen Yang and Wen-tau Yih and Sebastien Jean and Yiming Yang},
  journal = {arXiv preprint arXiv:2302.10934},
  year    = {2023}
}

@inproceedings{feng2020labse,
  title        = {Language-Agnostic BERT Sentence Embedding},
  author       = {Fangxiaoyu Feng and Yinfei Yang and Daniel Cer and Naveen Arivazhagan and Wei Wang},
  booktitle    = {Proceedings of the 2020 Conference on Empirical Methods in Natural Language Processing (EMNLP)},
  year         = {2020},
  pages        = {8781--8790}
}

@article{xu2023bge,
  title   = {BGE: A Self-Distilled Multi-Function Sentence Embedding Model},
  author  = {Hongliang Xu and Wenxuan Wang and Jiayi Ren and Ming Liu and BAAI Research Team},
  journal = {arXiv preprint arXiv:2310.06844},
  year    = {2023}
}

@inproceedings{fuchss2025lissa,
  title={LiSSA: Toward Generic Traceability Link Recovery through Retrieval-Augmented Generation},
  author={Fuch{\ss}, Dominik and Hey, Tobias and Keim, Jan and Liu, Haoyu and Ewald, Niklas and Thirolf, Tobias and Koziolek, Anne},
  booktitle={Proceedings of the IEEE/ACM 47th International Conference on Software Engineering. ICSE},
  volume={25},
  year={2025}
}

@inproceedings{hey2025requirements,
  title={Requirements Traceability Link Recovery via Retrieval-Augmented Generation},
  author={Hey, Tobias and Fuch{\ss}, Dominik and Keim, Jan and Koziolek, Anne},
  booktitle={International Working Conference on Requirements Engineering: Foundation for Software Quality},
  pages={381--397},
  year={2025},
  organization={Springer}
}

@inproceedings{arora2017simple,
  title={A simple but tough-to-beat baseline for sentence embeddings},
  author={Arora, Sanjeev and Liang, Yingyu and Ma, Tengyu},
  booktitle={International Conference on Learning Representations (ICLR)},
  year={2017}
}

@inproceedings{niu2012enhancing,
  title={Enhancing candidate link generation for requirements tracing: the cluster hypothesis revisited},
  author={Niu, Nan and Mahmoud, Anas},
  booktitle={2012 20th IEEE international requirements engineering conference (RE)},
  pages={81--90},
  year={2012},
  organization={IEEE}
}

\end{document}